\documentclass[aps,prl,twocolumn,superscriptaddress]{revtex4-1}

\usepackage{graphicx}
\usepackage{CJK}
\usepackage{dcolumn}
\usepackage{bm}
\usepackage{hyperref}
\usepackage{todonotes}
\presetkeys{todonotes}{inline,backgroundcolor=yellow}{}
    
\usepackage{color}
\usepackage{comment}
\usepackage{sidecap}
\usepackage{nicefrac}

\begin{document}

\title{Static Charge Density Wave Order in the Superconducting State of \LBCO{}}

\author{V. Thampy}
\email{vthampy@bnl.gov}
\altaffiliation[Present address: ]{Stanford Synchrotron Radiation Lightsource, SLAC National Accelerator Laboratory, CA 94025, USA}
\affiliation{Condensed Matter Physics and Materials Science Department, Brookhaven National Laboratory, Upton, New York 11973, USA}

\author{X. M. Chen}
\altaffiliation[Present address: ]{Advanced Light Source, Lawrence Berkeley National Laboratory, Berkeley, CA 94720, USA}
\affiliation{Condensed Matter Physics and Materials Science Department, Brookhaven National Laboratory, Upton, New York 11973, USA}

\author{Y. Cao}
\affiliation{Condensed Matter Physics and Materials Science Department, Brookhaven National Laboratory, Upton, New York 11973, USA}

\author{C. Mazzoli}
\author{A. M. Barbour}
\author{W. Hu}
\affiliation{National Synchrotron Light Source II, Brookhaven National Laboratory, Upton, New York 11973, USA}

\author{H. Miao}
\author{G. Fabbris}
\author{R. D. Zhong}
\author{G. D. Gu}
\author{J. M. Tranquada}
\author{I. K. Robinson}
\affiliation{Condensed Matter Physics and Materials Science Department, Brookhaven National Laboratory, Upton, New York 11973, USA}

\author{S. B. Wilkins}
\email{swilkins@bnl.gov }
\affiliation{National Synchrotron Light Source II, Brookhaven National Laboratory, Upton, New York 11973, USA}
\author{M. P. M. Dean}
\email{mdean@bnl.gov}
\affiliation{Condensed Matter Physics and Materials Science Department, Brookhaven National Laboratory, Upton, New York 11973, USA}

\def\mathbi#1{\ensuremath{\textbf{\em #1}}}
\def\Q{\ensuremath{\mathbi{Q}}}
\def\q{\ensuremath{\mathbi{q}}}
\def\LBCO{La$_{2-x}$Ba$_x$CuO$_4$}
\def\LSCO{La$_{2-x}$Sr$_x$CuO$_4$}
\def\LNSCO{La$_{2-x-y}$Nd$_y$Sr$_x$CuO$_4$}

\newcommand{\microns}{$\mathrm{\mu}$m}
\newcommand{\angstrom}{\mbox{\normalfont\AA}}
\date{\today}

\begin{abstract}
Charge density wave (CDW) correlations feature prominently in the phase diagram of the cuprates, motivating competing theories of whether fluctuating CDW correlations aid superconductivity or whether static CDW order coexists with superconductivity in inhomogeneous or spatially modulated states. Here we report Cu $L$-edge resonant x-ray photon correlation spectroscopy (XPCS) measurements of CDW correlations in superconducting \LBCO{} $x=0.11$. Static CDW order is shown to exist in the superconducting state at low temperatures and to persist up to at least 85\% of the CDW transition temperature. We discuss the implications of our observations for  how \emph{nominally} competing order parameters can coexist in the cuprates. 
\end{abstract}

%
\maketitle

\begin{figure}
    \includegraphics[width=0.45\textwidth]{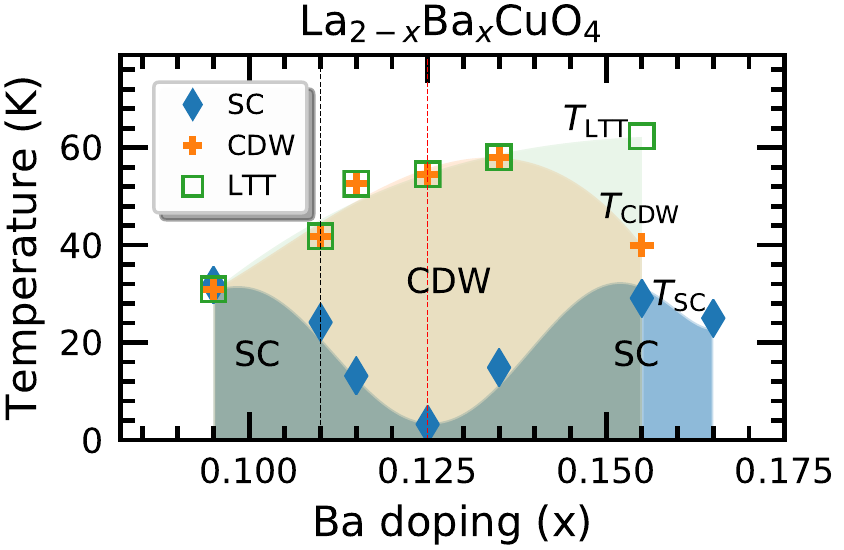}
    \caption{Doping phase diagram for \LBCO{} reproduced from Ref.~\cite{Hucker2011} showing regions with superconductivity (in blue) and how SC is suppressed around $x=0.125$ coincident with strongest CDW correlations. A vertical black dotted line marks the $x=0.11$ doping level studied here, which we compare with $x=0.125$ (red doted line).}
    \label{Intro}
\end{figure}

The properties of a complex material, such as a high temperature cuprate superconductor, are determined by its ground state configuration and spectrum of low energy fluctuations. For this reason, the discovery of charge density wave (CDW) correlations in various cuprates has attracted considerable attention \cite{Tranquada1995,Ghiringhelli2012,Chang2012,daSilvaNeto2014,Comin2014,Fujita2014,Thampy2014,Tabis2014}. Static charge order that develops from a conventional metallic state, as in various transition metal chalcongenides, tends to reduce the electronic density of states at the Fermi level and would be expected to suppress the superconducting transition temperature, $T_{\mathrm{SC}}$, of a standard BCS type superconductor \cite{Rossnagel2011}. Fluctuations  associated with incipient ordering tendencies, on the other hand, are often invoked in quantum critical theories of superconductivity \cite{Varma1997,Castellani1996,Sachdev1999,Vojta2000,Chakravarty2001,Sachdev2003,Abanov2003} and some theories posit CDW \cite{Castellani1996,Vojta2000,Wang2015,Caprara2016} or  nematic \cite{Maier2014,Lederer2015,Metlitski2015} fluctuations as the key modes. The motivation for such theories is clear in the context of the phase diagram of \LBCO{} plotted in Fig.~\ref{Intro}. At $x=0.125$ bulk superconductivity is almost completely suppressed coincident with the strongest CDW correlations \cite{Moodenbaugh1988, Hucker2011}, but at other dopings bulk superconductivity coexists with CDW correlations. 

We recently implemented Cu $L$-edge resonant x-ray photon correlation spectroscopy (XPCS) as a means to determine whether CDW correlations in the cuprates are static, fluctuating, or a combination of both \cite{Chen2016}. This technique has the advantage of a bulk probing depth, the ability to isolate the wavevector of charge (rather than spin) correlations and an excellent sensitivity to even very slow fluctuations.  Our first experiment showed that the CDW in \LBCO{} $x=0.125$ is robustly static up to 90\% of the CDW transition temperature, but this represents a special case in the LBCO phase diagram (Fig.\ref{Intro}) in which bulk superconductivity is almost completely suppressed \cite{Chen2016}.

In this Rapid Communication we demonstrate that static CDW correlations also exist in the superconducting state of \LBCO{} $x=0.11$ and that these static correlations extend above $T_{\mathrm{SC}}$ up to 40~K (85\% of the CDW transition temperature). To the extent that the superconducting mechanism in \LBCO{} is the same as in other cuprates, this sets important constraints on theories that suggest CDW fluctuations are crucial to superconductivity and how spatially-modulated states can be reconciled with superconductivity. 

\begin{figure}
    \includegraphics[width=0.45\textwidth]{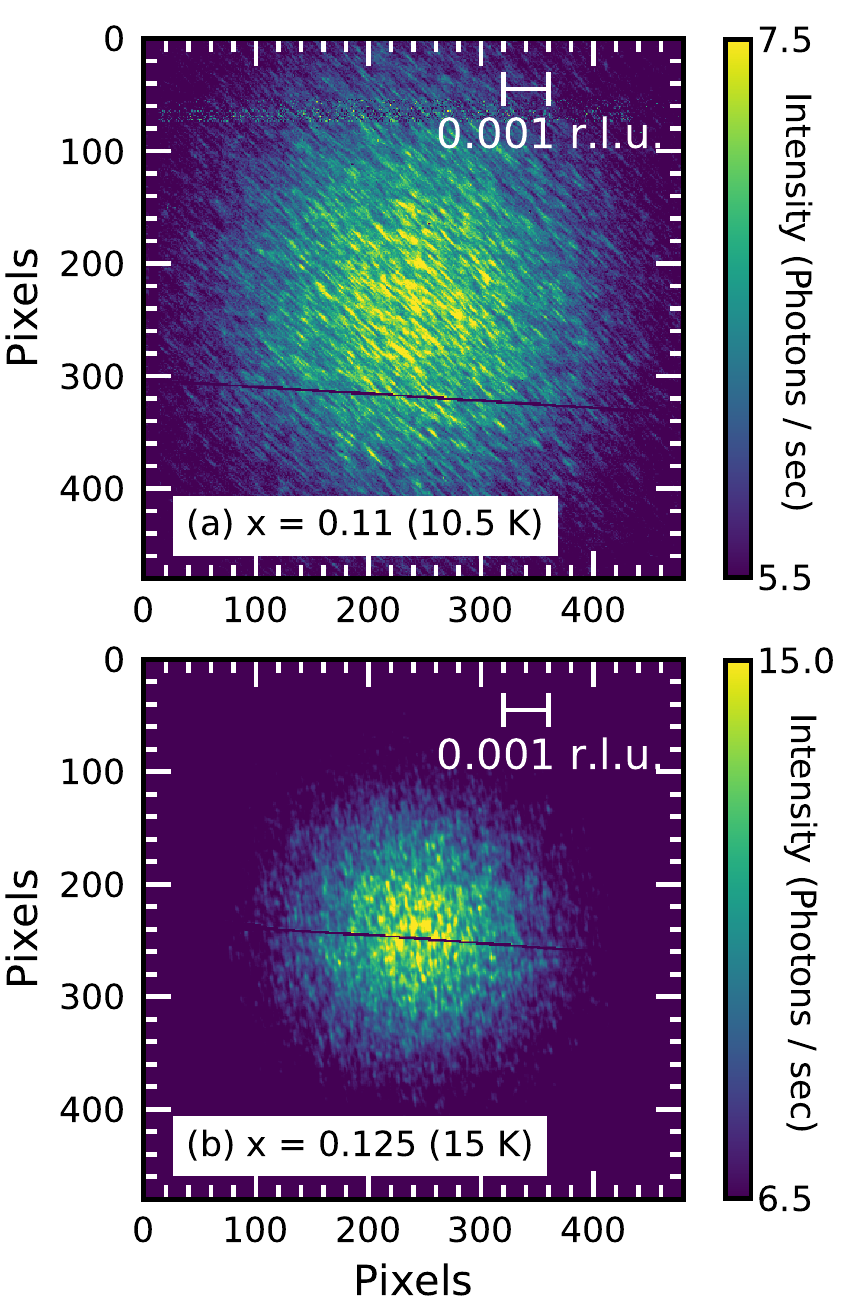}
    \caption{Speckle patterns for \LBCO{} with (a) $x=0.11$ and (b) $x=0.125$ \cite{Chen2016} taken at base temperatures of 10~K and 15~K respectively. The data are averaged over (a) 120 and (b) 165 minutes. We see that the CDW for $x=0.11$ has a significantly shorter correlation length of $150(8)$~\AA{} compared to $230(10)$~\AA{} for $x=0.125$. The horizontal direction in the images is approximately parallel to the $K$ direction in reciprocal space following the scale bars shown in white. Lines of reduced intensity are artifacts from the beamstop. }
    \label{speckles}
\end{figure}

\LBCO{} crystals were prepared using the floating zone method and cleaved \emph{ex-situ} to produce a scattering face with an approximate $c$-axis surface normal. Extensive previous studies on samples prepared in the same way demonstrated high sample quality with typical crystal mosaics of $\sim0.02^{\circ}$ \cite{Wilkins2011, DeanLBCO2013, Thampy2013, Dean2015, Chen2016, Miao2017}. XPCS experiments were performed at the 23-ID-1 beamline at the National Synchrotron Light Source II using a $[H,0,L]$ scattering plane as indexed in the high-temperature-tetragonal (HTT) unit cell with lattice constants $a=b=3.78$~\AA{} and $c=13.28$~\AA{}. In this notation, the CDW occurs at $(2\delta, 0, 0.5)$ where the incommensurability $\delta \approx x$ \cite{Fujita2004}. X-rays were focused onto a 10~\microns{} pinhole located $\sim$ 10~mm in front of the sample. Due to the diffraction-limited source and the beamline design, this leads to very high coherent flux ($\sim∼10^{13}$~photons/s) at the sample with a longitudinal coherence length of 2~\microns{} and a transverse coherence length of 10~\microns{} (set by the pinhole). A CCD \cite{fccd_camera} with a 30$\times$30~\microns{}$^2$ pixel size situated 340~mm from the sample was used to measure the CDW Bragg peak. All data were collected with horizontal ($\sigma$)-polarized incident x-rays at the Cu $L_3$-edge (931~eV) and an incident x-ray angle of approximately $33^{\circ}$.

\begin{figure}
    \includegraphics[width=0.45\textwidth]{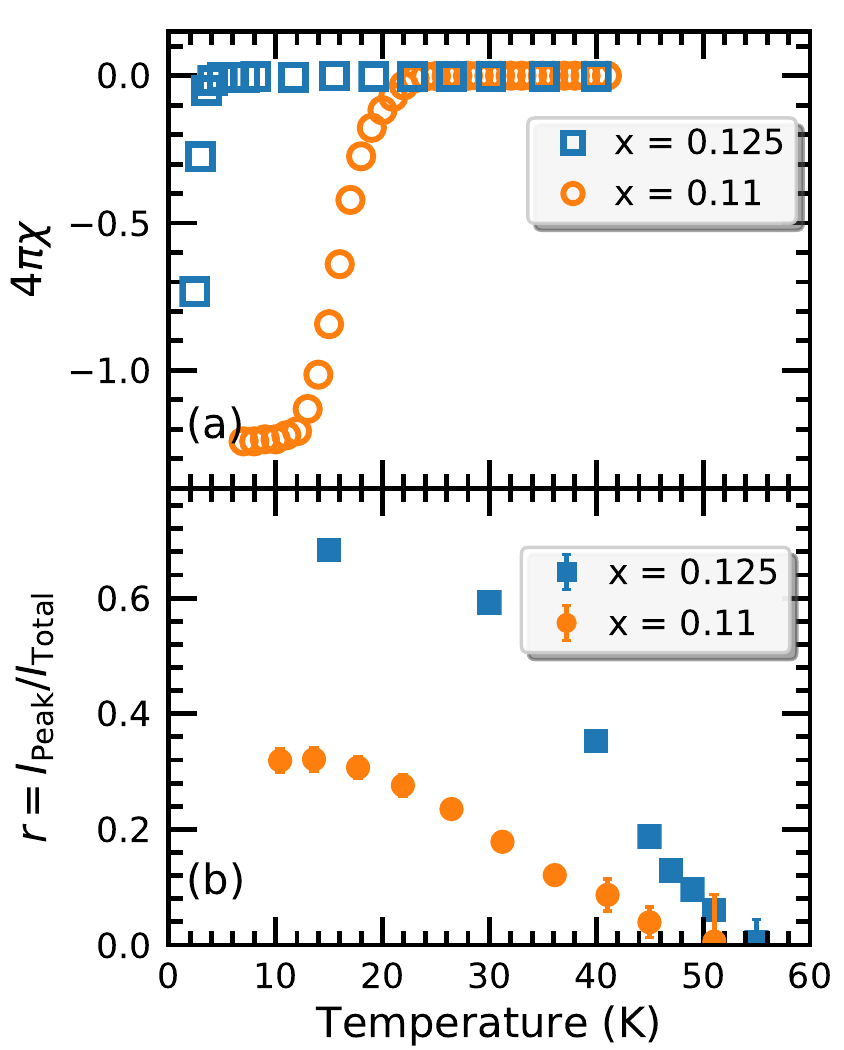}
    \caption{(a) Temperature dependence of the magnetic susceptibility measured in 1~mT applied field (after zero field cooling) for $x=0.11$ and $x=0.125$ samples. No demagnetization-correction has been applied. (b) The relative CDW peak intensity, $r$, defined as the intensity of the CDW scattering to the total (peak plus fluorescent) intensity. Errorbars on the $x=0.11$ data represent the sum of the standard deviation of the peak intensity parameter obtained by least-squares fitting and a 6\% uncertainty coming from variations in the scattering intensity from different spots on the sample, as estimated by comparing repeats of nominally equivalent measurements. The $x=0.125$ data is taken from Ref.~\cite{Chen2016} }
    \label{Tdep}
\end{figure}

Figure~\ref{speckles}(a) plots a detector image for \LBCO{} $x=0.11$ at the CDW wavevector $\mathbi{Q}=(-0.225, 0, 1.5)$ \cite{HKL_notation} summed over 120 minutes at 10~K. Speckle intensity modulations coming from coherent interference between different domains of CDW order appear throughout the image, modulating the overall peak shape. We compare this image to that obtained in \LBCO{} $x=0.125$ collected during previous measurements under similar conditions in Fig.~\ref{speckles}(b). $x=0.11$ shows a factor of 2 drop in the peak intensity on the detector relative to $x=0.125$ and a decrease in correlation length from 230 to 150~\AA{}, consistent with previous incoherent x-ray scattering results \cite{Hucker2011}. The elongation direction of the speckles is also different: being diagonal in Fig.~\ref{speckles}(a) and vertical Fig.~\ref{speckles}(b), which we assign to a deviation between the sample surface normal and the $c$-axis in the current $x=0.11$ sample \cite{Pitney2000}.

\begin{figure*}
    \includegraphics[width=\textwidth]{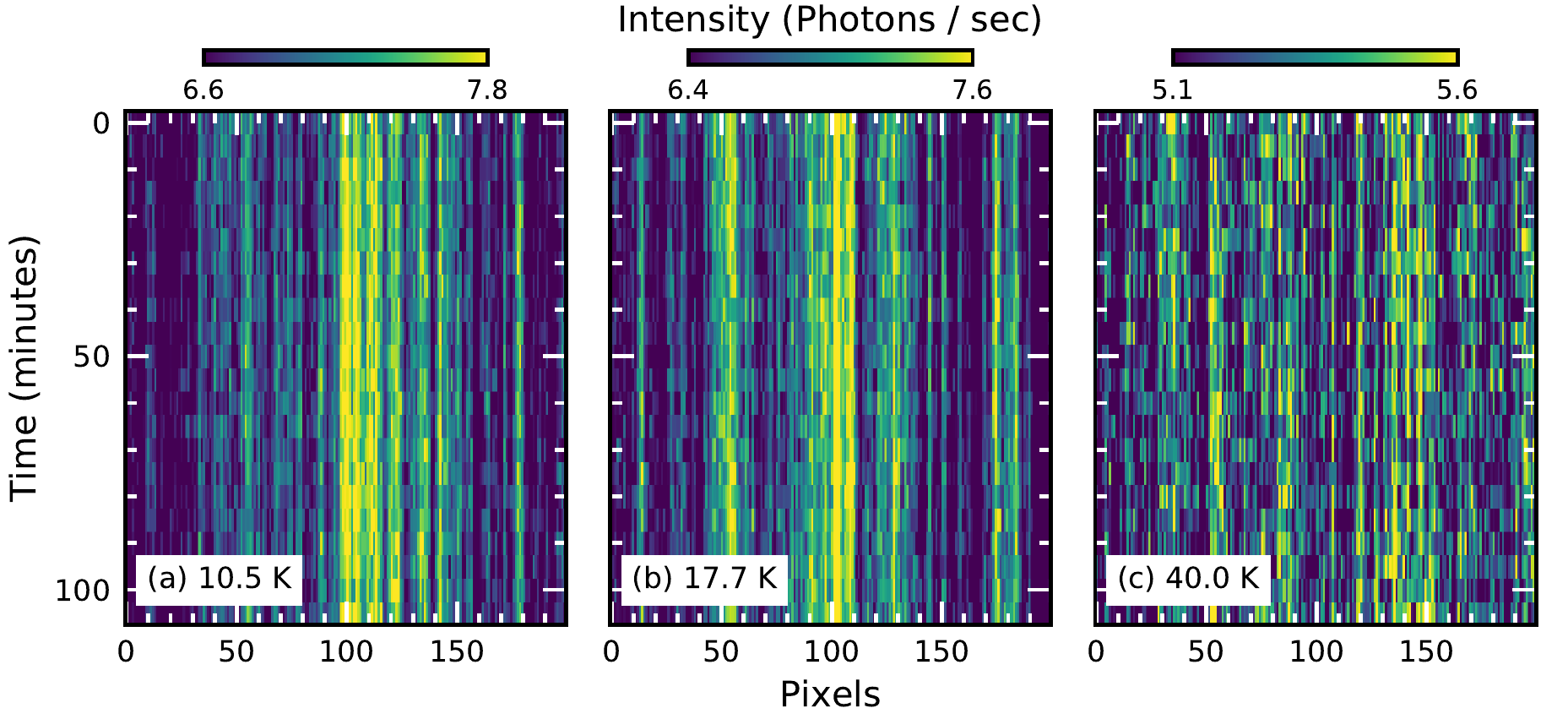}
    \caption{Waterfall plots showing the evolution of a line cut through the center of the CDW speckle pattern as a function of time at (a) 10.5~K, (b) 17.7~K and (c) 40.0~K. Vertical streaks are indicative of static CDW domains between 10.5-40.0~K. The vertical axis is binned into 5~minute time periods.}
    \label{Waterfalls}
\end{figure*}

Having measured the base temperature behavior, we explore possible variations with temperature. To provide context for this, we plot the magnetic susceptibility and relative CDW peak intensities at both doping levels in Fig.~\ref{Tdep}.
As seen in Fig.~\ref{Tdep}(a), \LBCO{} $x=0.125$ becomes  superconducting only below 5~K \cite{Li2007}, whereas for \LBCO{} $x=0.11$, the superconductivity occurs around 20~K with the magnetic susceptibility leveling off  below 11~K \cite{Hucker2011}. Figure~\ref{Tdep}(b) plots the relative CDW peak intensity $r$, defined as the ratio of the CDW scattering peak intensity to the total scattering intensity (peak + background). We choose this means of quantifying intensity as it will prove useful in evaluating the speckle visibility later in this work. The CDW order sets in at lower temperature for lower dopings: 48~K for $x=0.11$ versus 54~K for $x=0.125$ and grows in intensity as the temperature is reduced consistent with Ref.~\cite{Hucker2013}.


The time dependence of the CDW domain configurations was then tested at different temperatures by preparing ``waterfall'' plots, or kymographs, as shown in Fig.~\ref{Waterfalls}. Each panel is constructed by taking the same horizontal single-pixel-wide line-cut through the CDW peak at different times and stacking them on top of one another. Vertical lines in these plots come from speckles persisting at the same pixel and indicate static CDW domains. Despite the signal weakening with increasing temperature, these plots indicate static CDW behavior up to 40~K. We further quantified the statistical speckle behavior using the normalized one-time correlation function 
\begin{equation}
    g_2(\mathbi{q}, \tau) = \frac{\langle I(\mathbi{q},t)I(\mathbi{q}, t+\tau)\rangle}{\langle I(\mathbi{q}, t)\rangle ^2} = 1 + \beta \vert F(\mathbi{q},\tau)\vert ^2.
    \label{eq:g2}
\end{equation}
This correlates $I(\q,t)$, the intensity in a pixel at wavevector \q{} and time $t$, with the same quantity at lag time $\tau$ later \cite{Sutton2008, Shpyrko2014}. $\langle \dots \rangle$ indicates averaging over equivalent \q{} and different start times. $F(\q,\tau)$ is called the intermediate scattering function, which describes the time-dependent behavior of the sample and is defined as equal to 1 in the $\tau \rightarrow 0$ limit. Therefore, any reduction in $F(\q,\tau)$ at finite $\tau$ implies the presence of dynamics.  $\beta$ denotes the speckle contrast factor, the square of the optical visibility, which describes the magnitude of the speckle-modulated intensity 
\begin{equation}
    \beta = \left( \frac{I_{\text{s}}^{\text{max}} - I_{\text{s}}^{\text{min}}}{I_{\text{s}}^{\text{max}} + I_{\text{s}}^{\text{min}}}
    \right)^2 ,
\label{eq:I_speckle}
\end{equation}
where $I_{\text{s}}^{\text{max}}$ and $I_{\text{s}}^{\text{min}}$ are the maximum and minimum speckle-modulated intensities. Here we use the full measured intensity (from both the CDW and the fluorescence background) in order to avoid issues with subtracting background, which can be ambiguous when dealing with the weak CDW intensity at temperatures near the transition. $\beta$ can be reduced from its maximum value of 1 by the coherence properties of the incident beam, the background strength and intrinsic fast-timescale sample dynamics. As we go on to explain, the reduction in $\beta$ observed here can be assigned to the beam coherence and the signal-to-background ratio. In these measurements, the x-ray beam at the sample had horizontal and vertical transverse coherence lengths  of $\xi_{h}=\xi_{v}=10$~\microns{} (as set by the pinhole) and energy resolution $\Delta \lambda / \lambda \approx 1/1900$ giving a longitudinal coherence length $\xi_{\parallel} = 2.5$~\microns{}. Without taking into account the background coming primarily from x-ray fluorescence, the expected $\beta$ calculated based on the scattering angles and x-ray penetration depth is  $\beta \sim 0.04$ \footnote{The details on the calculations are available in the Supplementary Materials of Ref.~\cite{Chen2016}.}. A constant background further suppresses the speckle visibility by $r(T)^2$, where $r(T)$ was previously defined for Fig.~\ref{Tdep}(b) to yield a $\beta$ value of approximately 0.002 at our base temperature of 10~K. Empirically we measure $\beta= 0.003(1)$ consistent with all of the measured CDW intensity being static, otherwise we would expect to see reduced visibility.

\begin{figure}
    \includegraphics[width=0.45\textwidth]{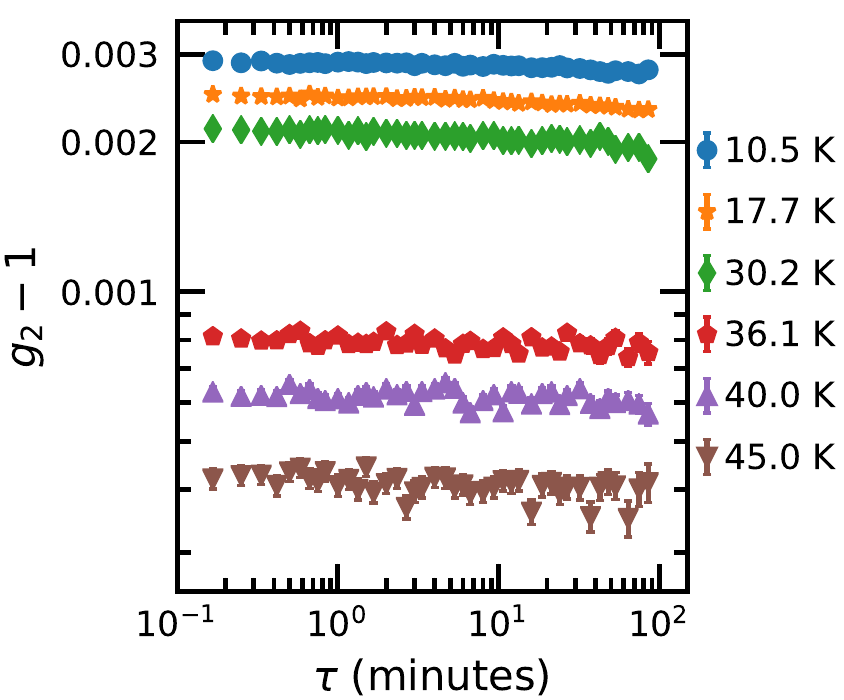}
    \caption{One-time correlation function, $g_2(\tau)-1$ for different sample temperatures. The curves are flat up to at least 100 minutes indicating static CDW correlations up to at least this timescale. We assign the small drop off in $g_2$ at longer timescales to finite beamline stability as explained in the text. Lower $g_2(\tau \rightarrow 0)$ values at high temperatures are consistent with reduced signal-to-background ratio, with no evidence for CDW fluctuations.}
    \label{g2}
\end{figure}

Figure~\ref{g2} plots $g_2(\tau)-1$ where we have explored low temperatures in the superconducting state, through the superconducting transition, up to temperatures approaching the CDW transition at 48~K. The base temperature data at 10.5~K is independent of timescale up to around $\tau \approx 100$~minutes.  As the sample is heated up, the $g_2(\tau)-1$ traces are roughly similar in shape, but with a reduction in $g_2(\tau \rightarrow 0) -1= \beta$, which comes from the drop in $r(T)$ near the transition \footnote{See the Supplemental Material at XXX for further details on the temperature scaling of the speckle contrast factor}. We consequently conclude that the CDW in  \LBCO{} $x=0.11$ is static up to at least 100~minutes up to a temperature of 45~K. At higher temperatures, the signal to noise ratio proved insufficient for a definitive statement.

These results show that static CDW order exists in the superconducting state, which has an onset temperature of 20~K and which saturates below 11~K. We believe this to be the first demonstration of static CDW order, on the timescale of hours, in the superconducting state of a `214'-type cuprate. We furthermore find that the speckle contrast is consistent with perfectly coherent scattering from the sample, with no evidence for significant fast fluctuations. Other techniques capable of detecting CDW order on similar macroscopic timescales to those studied here include scanning tunneling spectroscopy (STS), which has been applied to cleavable cuprates such as  Bi$_{2}$Sr$_{2}$Ca$_{n-1}$Cu$_{n}$O$_{2n+4+x}$ and Ca$_{2-x}$Na$_x$CuO$_2$Cl$_2$ \cite{Hoffman2002, Hanaguri2004, Parker2010, Comin2014, daSilvaNeto2015}. STS has demonstrated robustly static charge order, but is surface sensitive. Indeed, the difficulty of producing large, atomically flat surface areas has prevented STS imaging studies for various non-cleavable cuprates including \LBCO{} \cite{Valla2006}. CDWs in `214'-type cuprates are also highly challenging to study with nuclear magnetic resonance (NMR). Due to the lack of a spin gap, the signal tends to be dominated by spin fluctuations \cite{Julien2001, Pelc2017}. To date, many of the most compelling NMR studies have focused on YBa$_2$Cu$_3$O$_y$ \cite{Wu2011, Wu2013_NMR, Wu2015}, reporting static CDW order at low temperatures. It has been inferred that disorder-free YBa$_2$Cu$_3$O$_y$ \cite{Wu2015} might be dynamic, similar to classical CDW systems such as chromium \cite{Shpyrko2007} and TaS$_2$ \cite{Su2012}. 

How can we reconcile the existence of static CDW order below the superconducting transition in light of classical ideas that these two order parameters should compete? Trivial, macroscopic inhomogeneity is, a priori, possible, but as shown in Fig.~\ref{Tdep}(a) the magnetic susceptibility evinces a bulk superconductor similar to previously reported results \cite{Hucker2013}. Volume-sensitive probes of the CDW order parameter such as thermopower \cite{Tranquada2008}, muon spin rotation \cite{Savici2005} and nuclear magnetic/quadrapole resonance \cite{Hunt2001,Pelc2017}  also do not find evidence for significant inhomogeneity. More compellingly still, diffraction measurements show that CDW order is enhanced when superconductivity is suppressed by the application of magnetic field ruling out the idea that the ordering parameters are completely separated from one-another \cite{Hucker2013}.

One particular strength of XPCS is that it is sensitive to all the fluctuations within its measured range of wavevectors, in this case $\Delta Q \approx$~0.005~r.l.u.\ \cite{Sutton2008, Shpyrko2014}. Within this range, our data implies fully static CDW correlations on a timescale of hours and as we note above, we can exclude any significant fast fluctuations in terms of speckle visibility considerations. Our results therefore argue against any important role for CDW fluctuations around this wavevector for mediating superconductivity \cite{Castellani1996,Vojta2000,Wang2015,Caprara2016}. An interesting recent development is the discovery of finite charge correlations even at wavevectors very far from the central ordering wavevector, consistent with the idea that the CDW peak arises from a small fraction of pinned precursor correlations \cite{Miao2017, Bozin2015, Reznik2006}. A role for such precursor correlations remains possible, but this is beyond the scope of the current study as the intensity of such correlations is too low to be studied, despite the exceptionally high sensitivity of the 23-ID-1 setup.

A final consideration relates to pair-density-wave states \cite{Berg2007,Corboz2014,Lee2014,Fradkin2015}. In such a state, the superconducting pair wave function intertwines with the CDW to form a spatially modulated state that is commensurate with the CDW but with twice the period. STS has recently provided direct evidence for such a state in Bi$_{2}$Sr$_{2}$CaCu$_{2}$O$_{8+x}$ \cite{Hamidian2016}. Transport and neutron scattering measurements are also suggestive of such a state in \LBCO{} \cite{Li2007,Xu2014}. This state is also compatible with our observation of static coexistence of CDW and superconducting order parameters \cite{Berg2007,Corboz2014,Lee2014,Fradkin2015}.  

In conclusion, we have used Cu $L_3$ edge resonant XPCS to measure the nature of the CDW correlations in \LBCO{} $x=0.11$. We demonstrate that static CDW correlations exist over a timescale of at least 100 minutes within the superconducting state and persist up to at least 85\% of the CDW transition temperature. The apparent dichotomy of competing but coexisting order parameters can be possibly reconciled by invoking pair density wave states.

\begin{acknowledgments}
 We thank Emil Bozin and Marc-Henri Julien for insightful discussions. X-ray scattering measurements by H.M., M.P.M.D.\ and J.M.T.\ were  supported by the Center for Emergent Superconductivity, an Energy Frontier Research Center funded by the U.S. DOE, Office of Basic Energy Sciences. This research used resources at the 23-ID-1 beamline of the National Synchrotron Light Source II, a U.S.\ Department of Energy (DOE) Office of Science User Facility operated for the DOE Office of Science by Brookhaven National Laboratory under Contract No. DE-SC0012704.
\end{acknowledgments}

\bibliography{refs}

\begin{thebibliography}{62}%
\makeatletter
\providecommand \@ifxundefined [1]{%
 \@ifx{#1\undefined}
}%
\providecommand \@ifnum [1]{%
 \ifnum #1\expandafter \@firstoftwo
 \else \expandafter \@secondoftwo
 \fi
}%
\providecommand \@ifx [1]{%
 \ifx #1\expandafter \@firstoftwo
 \else \expandafter \@secondoftwo
 \fi
}%
\providecommand \natexlab [1]{#1}%
\providecommand \enquote  [1]{``#1''}%
\providecommand \bibnamefont  [1]{#1}%
\providecommand \bibfnamefont [1]{#1}%
\providecommand \citenamefont [1]{#1}%
\providecommand \href@noop [0]{\@secondoftwo}%
\providecommand \href [0]{\begingroup \@sanitize@url \@href}%
\providecommand \@href[1]{\@@startlink{#1}\@@href}%
\providecommand \@@href[1]{\endgroup#1\@@endlink}%
\providecommand \@sanitize@url [0]{\catcode `\\12\catcode `\$12\catcode
  `\&12\catcode `\#12\catcode `\^12\catcode `\_12\catcode `\%12\relax}%
\providecommand \@@startlink[1]{}%
\providecommand \@@endlink[0]{}%
\providecommand \url  [0]{\begingroup\@sanitize@url \@url }%
\providecommand \@url [1]{\endgroup\@href {#1}{\urlprefix }}%
\providecommand \urlprefix  [0]{URL }%
\providecommand \Eprint [0]{\href }%
\providecommand \doibase [0]{http://dx.doi.org/}%
\providecommand \selectlanguage [0]{\@gobble}%
\providecommand \bibinfo  [0]{\@secondoftwo}%
\providecommand \bibfield  [0]{\@secondoftwo}%
\providecommand \translation [1]{[#1]}%
\providecommand \BibitemOpen [0]{}%
\providecommand \bibitemStop [0]{}%
\providecommand \bibitemNoStop [0]{.\EOS\space}%
\providecommand \EOS [0]{\spacefactor3000\relax}%
\providecommand \BibitemShut  [1]{\csname bibitem#1\endcsname}%
\let\auto@bib@innerbib\@empty
\bibitem [{\citenamefont {{Tranquada}}\ \emph {et~al.}(1995)\citenamefont
  {{Tranquada}}, \citenamefont {{Sternlieb}}, \citenamefont {{Axe}},
  \citenamefont {{Nakamura}},\ and\ \citenamefont {{Uchida}}}]{Tranquada1995}%
  \BibitemOpen
  \bibfield  {author} {\bibinfo {author} {\bibfnamefont {J.~M.}\ \bibnamefont
  {{Tranquada}}}, \bibinfo {author} {\bibfnamefont {B.~J.}\ \bibnamefont
  {{Sternlieb}}}, \bibinfo {author} {\bibfnamefont {J.~D.}\ \bibnamefont
  {{Axe}}}, \bibinfo {author} {\bibfnamefont {Y.}~\bibnamefont {{Nakamura}}}, \
  and\ \bibinfo {author} {\bibfnamefont {S.}~\bibnamefont {{Uchida}}},\ }\href
  {\doibase 10.1038/375561a0} {\bibfield  {journal} {\bibinfo  {journal}
  {Nature}\ }\textbf {\bibinfo {volume} {375}},\ \bibinfo {pages} {561}
  (\bibinfo {year} {1995})}\BibitemShut {NoStop}%
\bibitem [{\citenamefont {Ghiringhelli}\ \emph {et~al.}(2012)\citenamefont
  {Ghiringhelli}, \citenamefont {Le~Tacon}, \citenamefont {Minola},
  \citenamefont {Blanco-Canosa}, \citenamefont {Mazzoli}, \citenamefont
  {Brookes}, \citenamefont {De~Luca}, \citenamefont {Frano}, \citenamefont
  {Hawthorn}, \citenamefont {He}, \citenamefont {Loew}, \citenamefont {Sala},
  \citenamefont {Peets}, \citenamefont {Salluzzo}, \citenamefont {Schierle},
  \citenamefont {Sutarto}, \citenamefont {Sawatzky}, \citenamefont {Weschke},
  \citenamefont {Keimer},\ and\ \citenamefont {Braicovich}}]{Ghiringhelli2012}%
  \BibitemOpen
  \bibfield  {author} {\bibinfo {author} {\bibfnamefont {G.}~\bibnamefont
  {Ghiringhelli}}, \bibinfo {author} {\bibfnamefont {M.}~\bibnamefont
  {Le~Tacon}}, \bibinfo {author} {\bibfnamefont {M.}~\bibnamefont {Minola}},
  \bibinfo {author} {\bibfnamefont {S.}~\bibnamefont {Blanco-Canosa}}, \bibinfo
  {author} {\bibfnamefont {C.}~\bibnamefont {Mazzoli}}, \bibinfo {author}
  {\bibfnamefont {N.~B.}\ \bibnamefont {Brookes}}, \bibinfo {author}
  {\bibfnamefont {G.~M.}\ \bibnamefont {De~Luca}}, \bibinfo {author}
  {\bibfnamefont {A.}~\bibnamefont {Frano}}, \bibinfo {author} {\bibfnamefont
  {D.~G.}\ \bibnamefont {Hawthorn}}, \bibinfo {author} {\bibfnamefont
  {F.}~\bibnamefont {He}}, \bibinfo {author} {\bibfnamefont {T.}~\bibnamefont
  {Loew}}, \bibinfo {author} {\bibfnamefont {M.~M.}\ \bibnamefont {Sala}},
  \bibinfo {author} {\bibfnamefont {D.~C.}\ \bibnamefont {Peets}}, \bibinfo
  {author} {\bibfnamefont {M.}~\bibnamefont {Salluzzo}}, \bibinfo {author}
  {\bibfnamefont {E.}~\bibnamefont {Schierle}}, \bibinfo {author}
  {\bibfnamefont {R.}~\bibnamefont {Sutarto}}, \bibinfo {author} {\bibfnamefont
  {G.~A.}\ \bibnamefont {Sawatzky}}, \bibinfo {author} {\bibfnamefont
  {E.}~\bibnamefont {Weschke}}, \bibinfo {author} {\bibfnamefont
  {B.}~\bibnamefont {Keimer}}, \ and\ \bibinfo {author} {\bibfnamefont
  {L.}~\bibnamefont {Braicovich}},\ }\href {\doibase 10.1126/science.1223532}
  {\bibfield  {journal} {\bibinfo  {journal} {Science}\ }\textbf {\bibinfo
  {volume} {337}},\ \bibinfo {pages} {821} (\bibinfo {year}
  {2012})}\BibitemShut {NoStop}%
\bibitem [{\citenamefont {{Chang}}\ \emph {et~al.}(2012)\citenamefont
  {{Chang}}, \citenamefont {{Blackburn}}, \citenamefont {{Holmes}},
  \citenamefont {{Christensen}}, \citenamefont {{Larsen}}, \citenamefont
  {{Mesot}}, \citenamefont {{Liang}}, \citenamefont {{Bonn}}, \citenamefont
  {{Hardy}}, \citenamefont {{Watenphul}}, \citenamefont {{Zimmermann}},
  \citenamefont {{Forgan}},\ and\ \citenamefont {{Hayden}}}]{Chang2012}%
  \BibitemOpen
  \bibfield  {author} {\bibinfo {author} {\bibfnamefont {J.}~\bibnamefont
  {{Chang}}}, \bibinfo {author} {\bibfnamefont {E.}~\bibnamefont
  {{Blackburn}}}, \bibinfo {author} {\bibfnamefont {A.~T.}\ \bibnamefont
  {{Holmes}}}, \bibinfo {author} {\bibfnamefont {N.~B.}\ \bibnamefont
  {{Christensen}}}, \bibinfo {author} {\bibfnamefont {J.}~\bibnamefont
  {{Larsen}}}, \bibinfo {author} {\bibfnamefont {J.}~\bibnamefont {{Mesot}}},
  \bibinfo {author} {\bibfnamefont {R.}~\bibnamefont {{Liang}}}, \bibinfo
  {author} {\bibfnamefont {D.~A.}\ \bibnamefont {{Bonn}}}, \bibinfo {author}
  {\bibfnamefont {W.~N.}\ \bibnamefont {{Hardy}}}, \bibinfo {author}
  {\bibfnamefont {A.}~\bibnamefont {{Watenphul}}}, \bibinfo {author}
  {\bibfnamefont {M.~V.}\ \bibnamefont {{Zimmermann}}}, \bibinfo {author}
  {\bibfnamefont {E.~M.}\ \bibnamefont {{Forgan}}}, \ and\ \bibinfo {author}
  {\bibfnamefont {S.~M.}\ \bibnamefont {{Hayden}}},\ }\href {\doibase
  10.1038/nphys2456} {\bibfield  {journal} {\bibinfo  {journal} {Nature
  Physics}\ }\textbf {\bibinfo {volume} {8}},\ \bibinfo {pages} {871} (\bibinfo
  {year} {2012})}\BibitemShut {NoStop}%
\bibitem [{\citenamefont {da~Silva~Neto}\ \emph {et~al.}(2014)\citenamefont
  {da~Silva~Neto}, \citenamefont {Aynajian}, \citenamefont {Frano},
  \citenamefont {Comin}, \citenamefont {Schierle}, \citenamefont {Weschke},
  \citenamefont {Gyenis}, \citenamefont {Wen}, \citenamefont {Schneeloch},
  \citenamefont {Xu} \emph {et~al.}}]{daSilvaNeto2014}%
  \BibitemOpen
  \bibfield  {author} {\bibinfo {author} {\bibfnamefont {E.~H.}\ \bibnamefont
  {da~Silva~Neto}}, \bibinfo {author} {\bibfnamefont {P.}~\bibnamefont
  {Aynajian}}, \bibinfo {author} {\bibfnamefont {A.}~\bibnamefont {Frano}},
  \bibinfo {author} {\bibfnamefont {R.}~\bibnamefont {Comin}}, \bibinfo
  {author} {\bibfnamefont {E.}~\bibnamefont {Schierle}}, \bibinfo {author}
  {\bibfnamefont {E.}~\bibnamefont {Weschke}}, \bibinfo {author} {\bibfnamefont
  {A.}~\bibnamefont {Gyenis}}, \bibinfo {author} {\bibfnamefont
  {J.}~\bibnamefont {Wen}}, \bibinfo {author} {\bibfnamefont {J.}~\bibnamefont
  {Schneeloch}}, \bibinfo {author} {\bibfnamefont {Z.}~\bibnamefont {Xu}},
  \emph {et~al.},\ }\href@noop {} {\bibfield  {journal} {\bibinfo  {journal}
  {Science}\ }\textbf {\bibinfo {volume} {343}},\ \bibinfo {pages} {393}
  (\bibinfo {year} {2014})}\BibitemShut {NoStop}%
\bibitem [{\citenamefont {Comin}\ \emph {et~al.}(2014)\citenamefont {Comin},
  \citenamefont {Frano}, \citenamefont {Yee}, \citenamefont {Yoshida},
  \citenamefont {Eisaki}, \citenamefont {Schierle}, \citenamefont {Weschke},
  \citenamefont {Sutarto}, \citenamefont {He}, \citenamefont {Soumyanarayanan},
  \citenamefont {He}, \citenamefont {Le~Tacon}, \citenamefont {Elfimov},
  \citenamefont {Hoffman}, \citenamefont {Sawatzky}, \citenamefont {Keimer},\
  and\ \citenamefont {Damascelli}}]{Comin2014}%
  \BibitemOpen
  \bibfield  {author} {\bibinfo {author} {\bibfnamefont {R.}~\bibnamefont
  {Comin}}, \bibinfo {author} {\bibfnamefont {A.}~\bibnamefont {Frano}},
  \bibinfo {author} {\bibfnamefont {M.~M.}\ \bibnamefont {Yee}}, \bibinfo
  {author} {\bibfnamefont {Y.}~\bibnamefont {Yoshida}}, \bibinfo {author}
  {\bibfnamefont {H.}~\bibnamefont {Eisaki}}, \bibinfo {author} {\bibfnamefont
  {E.}~\bibnamefont {Schierle}}, \bibinfo {author} {\bibfnamefont
  {E.}~\bibnamefont {Weschke}}, \bibinfo {author} {\bibfnamefont
  {R.}~\bibnamefont {Sutarto}}, \bibinfo {author} {\bibfnamefont
  {F.}~\bibnamefont {He}}, \bibinfo {author} {\bibfnamefont {A.}~\bibnamefont
  {Soumyanarayanan}}, \bibinfo {author} {\bibfnamefont {Y.}~\bibnamefont {He}},
  \bibinfo {author} {\bibfnamefont {M.}~\bibnamefont {Le~Tacon}}, \bibinfo
  {author} {\bibfnamefont {I.~S.}\ \bibnamefont {Elfimov}}, \bibinfo {author}
  {\bibfnamefont {J.~E.}\ \bibnamefont {Hoffman}}, \bibinfo {author}
  {\bibfnamefont {G.~A.}\ \bibnamefont {Sawatzky}}, \bibinfo {author}
  {\bibfnamefont {B.}~\bibnamefont {Keimer}}, \ and\ \bibinfo {author}
  {\bibfnamefont {A.}~\bibnamefont {Damascelli}},\ }\href {\doibase
  10.1126/science.1242996} {\bibfield  {journal} {\bibinfo  {journal}
  {Science}\ }\textbf {\bibinfo {volume} {343}},\ \bibinfo {pages} {390}
  (\bibinfo {year} {2014})}\BibitemShut {NoStop}%
\bibitem [{\citenamefont {Fujita}\ \emph {et~al.}(2014)\citenamefont {Fujita},
  \citenamefont {Hamidian}, \citenamefont {Edkins}, \citenamefont {Kim},
  \citenamefont {Kohsaka}, \citenamefont {Azuma}, \citenamefont {Takano},
  \citenamefont {Takagi}, \citenamefont {Eisaki}, \citenamefont {Uchida},
  \citenamefont {Allais}, \citenamefont {Lawler}, \citenamefont {Kim},
  \citenamefont {Sachdev},\ and\ \citenamefont {Davis}}]{Fujita2014}%
  \BibitemOpen
  \bibfield  {author} {\bibinfo {author} {\bibfnamefont {K.}~\bibnamefont
  {Fujita}}, \bibinfo {author} {\bibfnamefont {M.~H.}\ \bibnamefont
  {Hamidian}}, \bibinfo {author} {\bibfnamefont {S.~D.}\ \bibnamefont
  {Edkins}}, \bibinfo {author} {\bibfnamefont {C.~K.}\ \bibnamefont {Kim}},
  \bibinfo {author} {\bibfnamefont {Y.}~\bibnamefont {Kohsaka}}, \bibinfo
  {author} {\bibfnamefont {M.}~\bibnamefont {Azuma}}, \bibinfo {author}
  {\bibfnamefont {M.}~\bibnamefont {Takano}}, \bibinfo {author} {\bibfnamefont
  {H.}~\bibnamefont {Takagi}}, \bibinfo {author} {\bibfnamefont
  {H.}~\bibnamefont {Eisaki}}, \bibinfo {author} {\bibfnamefont {S.-i.}\
  \bibnamefont {Uchida}}, \bibinfo {author} {\bibfnamefont {A.}~\bibnamefont
  {Allais}}, \bibinfo {author} {\bibfnamefont {M.~J.}\ \bibnamefont {Lawler}},
  \bibinfo {author} {\bibfnamefont {E.-A.}\ \bibnamefont {Kim}}, \bibinfo
  {author} {\bibfnamefont {S.}~\bibnamefont {Sachdev}}, \ and\ \bibinfo
  {author} {\bibfnamefont {J.~C.~S.}\ \bibnamefont {Davis}},\ }\href {\doibase
  10.1073/pnas.1406297111} {\bibfield  {journal} {\bibinfo  {journal} {Proc.
  Natl. Acad. Sci. USA}\ }\textbf {\bibinfo {volume} {111}},\ \bibinfo {pages}
  {E3026} (\bibinfo {year} {2014})}\BibitemShut {NoStop}%
\bibitem [{\citenamefont {Thampy}\ \emph {et~al.}(2014)\citenamefont {Thampy},
  \citenamefont {Dean}, \citenamefont {Christensen}, \citenamefont {Steinke},
  \citenamefont {Islam}, \citenamefont {Oda}, \citenamefont {Ido},
  \citenamefont {Momono}, \citenamefont {Wilkins},\ and\ \citenamefont
  {Hill}}]{Thampy2014}%
  \BibitemOpen
  \bibfield  {author} {\bibinfo {author} {\bibfnamefont {V.}~\bibnamefont
  {Thampy}}, \bibinfo {author} {\bibfnamefont {M.~P.~M.}\ \bibnamefont {Dean}},
  \bibinfo {author} {\bibfnamefont {N.~B.}\ \bibnamefont {Christensen}},
  \bibinfo {author} {\bibfnamefont {L.}~\bibnamefont {Steinke}}, \bibinfo
  {author} {\bibfnamefont {Z.}~\bibnamefont {Islam}}, \bibinfo {author}
  {\bibfnamefont {M.}~\bibnamefont {Oda}}, \bibinfo {author} {\bibfnamefont
  {M.}~\bibnamefont {Ido}}, \bibinfo {author} {\bibfnamefont {N.}~\bibnamefont
  {Momono}}, \bibinfo {author} {\bibfnamefont {S.~B.}\ \bibnamefont {Wilkins}},
  \ and\ \bibinfo {author} {\bibfnamefont {J.~P.}\ \bibnamefont {Hill}},\
  }\href {\doibase 10.1103/PhysRevB.90.100510} {\bibfield  {journal} {\bibinfo
  {journal} {Phys. Rev. B}\ }\textbf {\bibinfo {volume} {90}},\ \bibinfo
  {pages} {100510} (\bibinfo {year} {2014})}\BibitemShut {NoStop}%
\bibitem [{\citenamefont {Tabis}\ \emph {et~al.}(2014)\citenamefont {Tabis},
  \citenamefont {Li}, \citenamefont {Le~Tacon}, \citenamefont {Braicovich},
  \citenamefont {Kreyssig}, \citenamefont {Minola}, \citenamefont {Dellea},
  \citenamefont {Weschke}, \citenamefont {Veit}, \citenamefont {Ramazanoglu}
  \emph {et~al.}}]{Tabis2014}%
  \BibitemOpen
  \bibfield  {author} {\bibinfo {author} {\bibfnamefont {W.}~\bibnamefont
  {Tabis}}, \bibinfo {author} {\bibfnamefont {Y.}~\bibnamefont {Li}}, \bibinfo
  {author} {\bibfnamefont {M.}~\bibnamefont {Le~Tacon}}, \bibinfo {author}
  {\bibfnamefont {L.}~\bibnamefont {Braicovich}}, \bibinfo {author}
  {\bibfnamefont {A.}~\bibnamefont {Kreyssig}}, \bibinfo {author}
  {\bibfnamefont {M.}~\bibnamefont {Minola}}, \bibinfo {author} {\bibfnamefont
  {G.}~\bibnamefont {Dellea}}, \bibinfo {author} {\bibfnamefont
  {E.}~\bibnamefont {Weschke}}, \bibinfo {author} {\bibfnamefont
  {M.}~\bibnamefont {Veit}}, \bibinfo {author} {\bibfnamefont {M.}~\bibnamefont
  {Ramazanoglu}},  \emph {et~al.},\ }\href@noop {} {\bibfield  {journal}
  {\bibinfo  {journal} {Nature communications}\ }\textbf {\bibinfo {volume}
  {5}} (\bibinfo {year} {2014})}\BibitemShut {NoStop}%
\bibitem [{\citenamefont {Rossnagel}(2011)}]{Rossnagel2011}%
  \BibitemOpen
  \bibfield  {author} {\bibinfo {author} {\bibfnamefont {K.}~\bibnamefont
  {Rossnagel}},\ }\href@noop {} {\bibfield  {journal} {\bibinfo  {journal}
  {Journal of Physics: Condensed Matter}\ }\textbf {\bibinfo {volume} {23}},\
  \bibinfo {pages} {213001} (\bibinfo {year} {2011})}\BibitemShut {NoStop}%
\bibitem [{\citenamefont {Varma}(1997)}]{Varma1997}%
  \BibitemOpen
  \bibfield  {author} {\bibinfo {author} {\bibfnamefont {C.~M.}\ \bibnamefont
  {Varma}},\ }\href {\doibase 10.1103/PhysRevB.55.14554} {\bibfield  {journal}
  {\bibinfo  {journal} {Phys. Rev. B}\ }\textbf {\bibinfo {volume} {55}},\
  \bibinfo {pages} {14554} (\bibinfo {year} {1997})}\BibitemShut {NoStop}%
\bibitem [{\citenamefont {Castellani}\ \emph {et~al.}(1996)\citenamefont
  {Castellani}, \citenamefont {Di~Castro},\ and\ \citenamefont
  {Grilli}}]{Castellani1996}%
  \BibitemOpen
  \bibfield  {author} {\bibinfo {author} {\bibfnamefont {C.}~\bibnamefont
  {Castellani}}, \bibinfo {author} {\bibfnamefont {C.}~\bibnamefont
  {Di~Castro}}, \ and\ \bibinfo {author} {\bibfnamefont {M.}~\bibnamefont
  {Grilli}},\ }\href {\doibase 10.1007/s002570050347} {\bibfield  {journal}
  {\bibinfo  {journal} {Zeitschrift f{\"u}r Physik B Condensed Matter}\
  }\textbf {\bibinfo {volume} {103}},\ \bibinfo {pages} {137} (\bibinfo {year}
  {1996})}\BibitemShut {NoStop}%
\bibitem [{\citenamefont {Sachdev}(1999)}]{Sachdev1999}%
  \BibitemOpen
  \bibfield  {author} {\bibinfo {author} {\bibfnamefont {S.}~\bibnamefont
  {Sachdev}},\ }\href {\doibase 10.1103/PhysRevB.59.14054} {\bibfield
  {journal} {\bibinfo  {journal} {Phys. Rev. B}\ }\textbf {\bibinfo {volume}
  {59}},\ \bibinfo {pages} {14054} (\bibinfo {year} {1999})}\BibitemShut
  {NoStop}%
\bibitem [{\citenamefont {Vojta}\ \emph {et~al.}(2000)\citenamefont {Vojta},
  \citenamefont {Zhang},\ and\ \citenamefont {Sachdev}}]{Vojta2000}%
  \BibitemOpen
  \bibfield  {author} {\bibinfo {author} {\bibfnamefont {M.}~\bibnamefont
  {Vojta}}, \bibinfo {author} {\bibfnamefont {Y.}~\bibnamefont {Zhang}}, \ and\
  \bibinfo {author} {\bibfnamefont {S.}~\bibnamefont {Sachdev}},\ }\href
  {\doibase 10.1103/PhysRevB.62.6721} {\bibfield  {journal} {\bibinfo
  {journal} {Phys. Rev. B}\ }\textbf {\bibinfo {volume} {62}},\ \bibinfo
  {pages} {6721} (\bibinfo {year} {2000})}\BibitemShut {NoStop}%
\bibitem [{\citenamefont {Chakravarty}\ \emph {et~al.}(2001)\citenamefont
  {Chakravarty}, \citenamefont {Laughlin}, \citenamefont {Morr},\ and\
  \citenamefont {Nayak}}]{Chakravarty2001}%
  \BibitemOpen
  \bibfield  {author} {\bibinfo {author} {\bibfnamefont {S.}~\bibnamefont
  {Chakravarty}}, \bibinfo {author} {\bibfnamefont {R.~B.}\ \bibnamefont
  {Laughlin}}, \bibinfo {author} {\bibfnamefont {D.~K.}\ \bibnamefont {Morr}},
  \ and\ \bibinfo {author} {\bibfnamefont {C.}~\bibnamefont {Nayak}},\ }\href
  {\doibase 10.1103/PhysRevB.63.094503} {\bibfield  {journal} {\bibinfo
  {journal} {Phys. Rev. B}\ }\textbf {\bibinfo {volume} {63}},\ \bibinfo
  {pages} {094503} (\bibinfo {year} {2001})}\BibitemShut {NoStop}%
\bibitem [{\citenamefont {Sachdev}(2003)}]{Sachdev2003}%
  \BibitemOpen
  \bibfield  {author} {\bibinfo {author} {\bibfnamefont {S.}~\bibnamefont
  {Sachdev}},\ }\href {\doibase 10.1103/RevModPhys.75.913} {\bibfield
  {journal} {\bibinfo  {journal} {Rev. Mod. Phys.}\ }\textbf {\bibinfo {volume}
  {75}},\ \bibinfo {pages} {913} (\bibinfo {year} {2003})}\BibitemShut
  {NoStop}%
\bibitem [{\citenamefont {Abanov}\ \emph {et~al.}(2003)\citenamefont {Abanov},
  \citenamefont {Chubukov},\ and\ \citenamefont {Schmalian}}]{Abanov2003}%
  \BibitemOpen
  \bibfield  {author} {\bibinfo {author} {\bibfnamefont {A.}~\bibnamefont
  {Abanov}}, \bibinfo {author} {\bibfnamefont {A.~V.}\ \bibnamefont
  {Chubukov}}, \ and\ \bibinfo {author} {\bibfnamefont {J.}~\bibnamefont
  {Schmalian}},\ }\href {\doibase 10.1080/0001873021000057123} {\bibfield
  {journal} {\bibinfo  {journal} {Advances in Physics}\ }\textbf {\bibinfo
  {volume} {52}},\ \bibinfo {pages} {119} (\bibinfo {year} {2003})},\ \Eprint
  {http://arxiv.org/abs/http://dx.doi.org/10.1080/0001873021000057123}
  {http://dx.doi.org/10.1080/0001873021000057123} \BibitemShut {NoStop}%
\bibitem [{\citenamefont {Wang}\ and\ \citenamefont
  {Chubukov}(2015)}]{Wang2015}%
  \BibitemOpen
  \bibfield  {author} {\bibinfo {author} {\bibfnamefont {Y.}~\bibnamefont
  {Wang}}\ and\ \bibinfo {author} {\bibfnamefont {A.~V.}\ \bibnamefont
  {Chubukov}},\ }\href {\doibase 10.1103/PhysRevB.92.125108} {\bibfield
  {journal} {\bibinfo  {journal} {Phys. Rev. B}\ }\textbf {\bibinfo {volume}
  {92}},\ \bibinfo {pages} {125108} (\bibinfo {year} {2015})}\BibitemShut
  {NoStop}%
\bibitem [{\citenamefont {{Caprara}}\ \emph {et~al.}(2016)\citenamefont
  {{Caprara}}, \citenamefont {{Di Castro}}, \citenamefont {{Seibold}},\ and\
  \citenamefont {{Grilli}}}]{Caprara2016}%
  \BibitemOpen
  \bibfield  {author} {\bibinfo {author} {\bibfnamefont {S.}~\bibnamefont
  {{Caprara}}}, \bibinfo {author} {\bibfnamefont {C.}~\bibnamefont {{Di
  Castro}}}, \bibinfo {author} {\bibfnamefont {G.}~\bibnamefont {{Seibold}}}, \
  and\ \bibinfo {author} {\bibfnamefont {M.}~\bibnamefont {{Grilli}}},\
  }\href@noop {} {\bibfield  {journal} {\bibinfo  {journal} {ArXiv e-prints}\ }
  (\bibinfo {year} {2016})},\ \bibinfo {note} {(Accepted in Phys. Rev. B)},\
  \Eprint {http://arxiv.org/abs/1604.07852} {arXiv:1604.07852
  [cond-mat.supr-con]} \BibitemShut {NoStop}%
\bibitem [{\citenamefont {Maier}\ and\ \citenamefont
  {Scalapino}(2014)}]{Maier2014}%
  \BibitemOpen
  \bibfield  {author} {\bibinfo {author} {\bibfnamefont {T.~A.}\ \bibnamefont
  {Maier}}\ and\ \bibinfo {author} {\bibfnamefont {D.~J.}\ \bibnamefont
  {Scalapino}},\ }\href {\doibase 10.1103/PhysRevB.90.174510} {\bibfield
  {journal} {\bibinfo  {journal} {Phys. Rev. B}\ }\textbf {\bibinfo {volume}
  {90}},\ \bibinfo {pages} {174510} (\bibinfo {year} {2014})}\BibitemShut
  {NoStop}%
\bibitem [{\citenamefont {Lederer}\ \emph {et~al.}(2015)\citenamefont
  {Lederer}, \citenamefont {Schattner}, \citenamefont {Berg},\ and\
  \citenamefont {Kivelson}}]{Lederer2015}%
  \BibitemOpen
  \bibfield  {author} {\bibinfo {author} {\bibfnamefont {S.}~\bibnamefont
  {Lederer}}, \bibinfo {author} {\bibfnamefont {Y.}~\bibnamefont {Schattner}},
  \bibinfo {author} {\bibfnamefont {E.}~\bibnamefont {Berg}}, \ and\ \bibinfo
  {author} {\bibfnamefont {S.~A.}\ \bibnamefont {Kivelson}},\ }\href {\doibase
  10.1103/PhysRevLett.114.097001} {\bibfield  {journal} {\bibinfo  {journal}
  {Phys. Rev. Lett.}\ }\textbf {\bibinfo {volume} {114}},\ \bibinfo {pages}
  {097001} (\bibinfo {year} {2015})}\BibitemShut {NoStop}%
\bibitem [{\citenamefont {Metlitski}\ \emph {et~al.}(2015)\citenamefont
  {Metlitski}, \citenamefont {Mross}, \citenamefont {Sachdev},\ and\
  \citenamefont {Senthil}}]{Metlitski2015}%
  \BibitemOpen
  \bibfield  {author} {\bibinfo {author} {\bibfnamefont {M.~A.}\ \bibnamefont
  {Metlitski}}, \bibinfo {author} {\bibfnamefont {D.~F.}\ \bibnamefont
  {Mross}}, \bibinfo {author} {\bibfnamefont {S.}~\bibnamefont {Sachdev}}, \
  and\ \bibinfo {author} {\bibfnamefont {T.}~\bibnamefont {Senthil}},\ }\href
  {\doibase 10.1103/PhysRevB.91.115111} {\bibfield  {journal} {\bibinfo
  {journal} {Phys. Rev. B}\ }\textbf {\bibinfo {volume} {91}},\ \bibinfo
  {pages} {115111} (\bibinfo {year} {2015})}\BibitemShut {NoStop}%
\bibitem [{\citenamefont {Moodenbaugh}\ \emph {et~al.}(1988)\citenamefont
  {Moodenbaugh}, \citenamefont {Xu}, \citenamefont {Suenaga}, \citenamefont
  {Folkerts},\ and\ \citenamefont {Shelton}}]{Moodenbaugh1988}%
  \BibitemOpen
  \bibfield  {author} {\bibinfo {author} {\bibfnamefont {A.~R.}\ \bibnamefont
  {Moodenbaugh}}, \bibinfo {author} {\bibfnamefont {Y.}~\bibnamefont {Xu}},
  \bibinfo {author} {\bibfnamefont {M.}~\bibnamefont {Suenaga}}, \bibinfo
  {author} {\bibfnamefont {T.~J.}\ \bibnamefont {Folkerts}}, \ and\ \bibinfo
  {author} {\bibfnamefont {R.~N.}\ \bibnamefont {Shelton}},\ }\href {\doibase
  10.1103/PhysRevB.38.4596} {\bibfield  {journal} {\bibinfo  {journal} {Phys.
  Rev. B}\ }\textbf {\bibinfo {volume} {38}},\ \bibinfo {pages} {4596}
  (\bibinfo {year} {1988})}\BibitemShut {NoStop}%
\bibitem [{\citenamefont {H\"ucker}\ \emph {et~al.}(2011)\citenamefont
  {H\"ucker}, \citenamefont {v.~Zimmermann}, \citenamefont {Gu}, \citenamefont
  {Xu}, \citenamefont {Wen}, \citenamefont {Xu}, \citenamefont {Kang},
  \citenamefont {Zheludev},\ and\ \citenamefont {Tranquada}}]{Hucker2011}%
  \BibitemOpen
  \bibfield  {author} {\bibinfo {author} {\bibfnamefont {M.}~\bibnamefont
  {H\"ucker}}, \bibinfo {author} {\bibfnamefont {M.}~\bibnamefont
  {v.~Zimmermann}}, \bibinfo {author} {\bibfnamefont {G.~D.}\ \bibnamefont
  {Gu}}, \bibinfo {author} {\bibfnamefont {Z.~J.}\ \bibnamefont {Xu}}, \bibinfo
  {author} {\bibfnamefont {J.~S.}\ \bibnamefont {Wen}}, \bibinfo {author}
  {\bibfnamefont {G.}~\bibnamefont {Xu}}, \bibinfo {author} {\bibfnamefont
  {H.~J.}\ \bibnamefont {Kang}}, \bibinfo {author} {\bibfnamefont
  {A.}~\bibnamefont {Zheludev}}, \ and\ \bibinfo {author} {\bibfnamefont
  {J.~M.}\ \bibnamefont {Tranquada}},\ }\href {\doibase
  10.1103/PhysRevB.83.104506} {\bibfield  {journal} {\bibinfo  {journal} {Phys.
  Rev. B}\ }\textbf {\bibinfo {volume} {83}},\ \bibinfo {pages} {104506}
  (\bibinfo {year} {2011})}\BibitemShut {NoStop}%
\bibitem [{\citenamefont {Chen}\ \emph {et~al.}(2016)\citenamefont {Chen},
  \citenamefont {Thampy}, \citenamefont {Mazzoli}, \citenamefont {Barbour},
  \citenamefont {Miao}, \citenamefont {Gu}, \citenamefont {Cao}, \citenamefont
  {Tranquada}, \citenamefont {Dean},\ and\ \citenamefont {Wilkins}}]{Chen2016}%
  \BibitemOpen
  \bibfield  {author} {\bibinfo {author} {\bibfnamefont {X.~M.}\ \bibnamefont
  {Chen}}, \bibinfo {author} {\bibfnamefont {V.}~\bibnamefont {Thampy}},
  \bibinfo {author} {\bibfnamefont {C.}~\bibnamefont {Mazzoli}}, \bibinfo
  {author} {\bibfnamefont {A.~M.}\ \bibnamefont {Barbour}}, \bibinfo {author}
  {\bibfnamefont {H.}~\bibnamefont {Miao}}, \bibinfo {author} {\bibfnamefont
  {G.~D.}\ \bibnamefont {Gu}}, \bibinfo {author} {\bibfnamefont
  {Y.}~\bibnamefont {Cao}}, \bibinfo {author} {\bibfnamefont {J.~M.}\
  \bibnamefont {Tranquada}}, \bibinfo {author} {\bibfnamefont {M.~P.~M.}\
  \bibnamefont {Dean}}, \ and\ \bibinfo {author} {\bibfnamefont {S.~B.}\
  \bibnamefont {Wilkins}},\ }\href {\doibase 10.1103/PhysRevLett.117.167001}
  {\bibfield  {journal} {\bibinfo  {journal} {Phys. Rev. Lett.}\ }\textbf
  {\bibinfo {volume} {117}},\ \bibinfo {pages} {167001} (\bibinfo {year}
  {2016})}\BibitemShut {NoStop}%
\bibitem [{\citenamefont {Wilkins}\ \emph {et~al.}(2011)\citenamefont
  {Wilkins}, \citenamefont {Dean}, \citenamefont {Fink}, \citenamefont
  {H\"ucker}, \citenamefont {Geck}, \citenamefont {Soltwisch}, \citenamefont
  {Schierle}, \citenamefont {Weschke}, \citenamefont {Gu}, \citenamefont
  {Uchida}, \citenamefont {Ichikawa}, \citenamefont {Tranquada},\ and\
  \citenamefont {Hill}}]{Wilkins2011}%
  \BibitemOpen
  \bibfield  {author} {\bibinfo {author} {\bibfnamefont {S.~B.}\ \bibnamefont
  {Wilkins}}, \bibinfo {author} {\bibfnamefont {M.~P.~M.}\ \bibnamefont
  {Dean}}, \bibinfo {author} {\bibfnamefont {J.}~\bibnamefont {Fink}}, \bibinfo
  {author} {\bibfnamefont {M.}~\bibnamefont {H\"ucker}}, \bibinfo {author}
  {\bibfnamefont {J.}~\bibnamefont {Geck}}, \bibinfo {author} {\bibfnamefont
  {V.}~\bibnamefont {Soltwisch}}, \bibinfo {author} {\bibfnamefont
  {E.}~\bibnamefont {Schierle}}, \bibinfo {author} {\bibfnamefont
  {E.}~\bibnamefont {Weschke}}, \bibinfo {author} {\bibfnamefont
  {G.}~\bibnamefont {Gu}}, \bibinfo {author} {\bibfnamefont {S.}~\bibnamefont
  {Uchida}}, \bibinfo {author} {\bibfnamefont {N.}~\bibnamefont {Ichikawa}},
  \bibinfo {author} {\bibfnamefont {J.~M.}\ \bibnamefont {Tranquada}}, \ and\
  \bibinfo {author} {\bibfnamefont {J.~P.}\ \bibnamefont {Hill}},\ }\href
  {\doibase 10.1103/PhysRevB.84.195101} {\bibfield  {journal} {\bibinfo
  {journal} {Phys. Rev. B}\ }\textbf {\bibinfo {volume} {84}},\ \bibinfo
  {pages} {195101} (\bibinfo {year} {2011})}\BibitemShut {NoStop}%
\bibitem [{\citenamefont {Dean}\ \emph {et~al.}(2013)\citenamefont {Dean},
  \citenamefont {Dellea}, \citenamefont {Minola}, \citenamefont {Wilkins},
  \citenamefont {Konik}, \citenamefont {Gu}, \citenamefont {Le~Tacon},
  \citenamefont {Brookes}, \citenamefont {Yakhou-Harris}, \citenamefont
  {Kummer}, \citenamefont {Hill}, \citenamefont {Braicovich},\ and\
  \citenamefont {Ghiringhelli}}]{DeanLBCO2013}%
  \BibitemOpen
  \bibfield  {author} {\bibinfo {author} {\bibfnamefont {M.~P.~M.}\
  \bibnamefont {Dean}}, \bibinfo {author} {\bibfnamefont {G.}~\bibnamefont
  {Dellea}}, \bibinfo {author} {\bibfnamefont {M.}~\bibnamefont {Minola}},
  \bibinfo {author} {\bibfnamefont {S.~B.}\ \bibnamefont {Wilkins}}, \bibinfo
  {author} {\bibfnamefont {R.~M.}\ \bibnamefont {Konik}}, \bibinfo {author}
  {\bibfnamefont {G.~D.}\ \bibnamefont {Gu}}, \bibinfo {author} {\bibfnamefont
  {M.}~\bibnamefont {Le~Tacon}}, \bibinfo {author} {\bibfnamefont {N.~B.}\
  \bibnamefont {Brookes}}, \bibinfo {author} {\bibfnamefont {F.}~\bibnamefont
  {Yakhou-Harris}}, \bibinfo {author} {\bibfnamefont {K.}~\bibnamefont
  {Kummer}}, \bibinfo {author} {\bibfnamefont {J.~P.}\ \bibnamefont {Hill}},
  \bibinfo {author} {\bibfnamefont {L.}~\bibnamefont {Braicovich}}, \ and\
  \bibinfo {author} {\bibfnamefont {G.}~\bibnamefont {Ghiringhelli}},\ }\href
  {\doibase 10.1103/PhysRevB.88.020403} {\bibfield  {journal} {\bibinfo
  {journal} {Phys. Rev. B}\ }\textbf {\bibinfo {volume} {88}},\ \bibinfo
  {pages} {020403} (\bibinfo {year} {2013})}\BibitemShut {NoStop}%
\bibitem [{\citenamefont {Thampy}\ \emph {et~al.}(2013)\citenamefont {Thampy},
  \citenamefont {Blanco-Canosa}, \citenamefont {Garcia-Fernandez},
  \citenamefont {Dean}, \citenamefont {Gu}, \citenamefont {F\"orst},
  \citenamefont {Loew}, \citenamefont {Keimer}, \citenamefont {Le~Tacon},
  \citenamefont {Wilkins},\ and\ \citenamefont {Hill}}]{Thampy2013}%
  \BibitemOpen
  \bibfield  {author} {\bibinfo {author} {\bibfnamefont {V.}~\bibnamefont
  {Thampy}}, \bibinfo {author} {\bibfnamefont {S.}~\bibnamefont
  {Blanco-Canosa}}, \bibinfo {author} {\bibfnamefont {M.}~\bibnamefont
  {Garcia-Fernandez}}, \bibinfo {author} {\bibfnamefont {M.~P.~M.}\
  \bibnamefont {Dean}}, \bibinfo {author} {\bibfnamefont {G.~D.}\ \bibnamefont
  {Gu}}, \bibinfo {author} {\bibfnamefont {M.}~\bibnamefont {F\"orst}},
  \bibinfo {author} {\bibfnamefont {T.}~\bibnamefont {Loew}}, \bibinfo {author}
  {\bibfnamefont {B.}~\bibnamefont {Keimer}}, \bibinfo {author} {\bibfnamefont
  {M.}~\bibnamefont {Le~Tacon}}, \bibinfo {author} {\bibfnamefont {S.~B.}\
  \bibnamefont {Wilkins}}, \ and\ \bibinfo {author} {\bibfnamefont {J.~P.}\
  \bibnamefont {Hill}},\ }\href {\doibase 10.1103/PhysRevB.88.024505}
  {\bibfield  {journal} {\bibinfo  {journal} {Phys. Rev. B}\ }\textbf {\bibinfo
  {volume} {88}},\ \bibinfo {pages} {024505} (\bibinfo {year}
  {2013})}\BibitemShut {NoStop}%
\bibitem [{\citenamefont {Dean}(2015)}]{Dean2015}%
  \BibitemOpen
  \bibfield  {author} {\bibinfo {author} {\bibfnamefont {M.~P.~M.}\
  \bibnamefont {Dean}},\ }\href {\doibase
  http://dx.doi.org/10.1016/j.jmmm.2014.03.057} {\bibfield  {journal} {\bibinfo
   {journal} {Journal of Magnetism and Magnetic Materials}\ }\textbf {\bibinfo
  {volume} {376}},\ \bibinfo {pages} {3 } (\bibinfo {year} {2015})}\BibitemShut
  {NoStop}%
\bibitem [{\citenamefont {{Miao}}\ \emph {et~al.}(2017)\citenamefont {{Miao}},
  \citenamefont {{Lorenzana}}, \citenamefont {{Seibold}}, \citenamefont
  {{Peng}}, \citenamefont {{Amorese}}, \citenamefont {{Yakhou-Harris}},
  \citenamefont {{Kummer}}, \citenamefont {{Brookes}}, \citenamefont {{Konik}},
  \citenamefont {{Thampy}}, \citenamefont {{Gu}}, \citenamefont
  {{Ghiringhelli}}, \citenamefont {{Braicovich}},\ and\ \citenamefont
  {{Dean}}}]{Miao2017}%
  \BibitemOpen
  \bibfield  {author} {\bibinfo {author} {\bibfnamefont {H.}~\bibnamefont
  {{Miao}}}, \bibinfo {author} {\bibfnamefont {J.}~\bibnamefont {{Lorenzana}}},
  \bibinfo {author} {\bibfnamefont {G.}~\bibnamefont {{Seibold}}}, \bibinfo
  {author} {\bibfnamefont {Y.~Y.}\ \bibnamefont {{Peng}}}, \bibinfo {author}
  {\bibfnamefont {A.}~\bibnamefont {{Amorese}}}, \bibinfo {author}
  {\bibfnamefont {F.}~\bibnamefont {{Yakhou-Harris}}}, \bibinfo {author}
  {\bibfnamefont {K.}~\bibnamefont {{Kummer}}}, \bibinfo {author}
  {\bibfnamefont {N.~B.}\ \bibnamefont {{Brookes}}}, \bibinfo {author}
  {\bibfnamefont {R.~M.}\ \bibnamefont {{Konik}}}, \bibinfo {author}
  {\bibfnamefont {V.}~\bibnamefont {{Thampy}}}, \bibinfo {author}
  {\bibfnamefont {G.~D.}\ \bibnamefont {{Gu}}}, \bibinfo {author}
  {\bibfnamefont {G.}~\bibnamefont {{Ghiringhelli}}}, \bibinfo {author}
  {\bibfnamefont {L.}~\bibnamefont {{Braicovich}}}, \ and\ \bibinfo {author}
  {\bibfnamefont {M.~P.~M.}\ \bibnamefont {{Dean}}},\ }\href@noop {} {\bibfield
   {journal} {\bibinfo  {journal} {ArXiv e-prints}\ } (\bibinfo {year}
  {2017})},\ \Eprint {http://arxiv.org/abs/1701.00022} {arXiv:1701.00022
  [cond-mat.supr-con]} \BibitemShut {NoStop}%
\bibitem [{\citenamefont {Fujita}\ \emph {et~al.}(2004)\citenamefont {Fujita},
  \citenamefont {Goka}, \citenamefont {Yamada}, \citenamefont {Tranquada},\
  and\ \citenamefont {Regnault}}]{Fujita2004}%
  \BibitemOpen
  \bibfield  {author} {\bibinfo {author} {\bibfnamefont {M.}~\bibnamefont
  {Fujita}}, \bibinfo {author} {\bibfnamefont {H.}~\bibnamefont {Goka}},
  \bibinfo {author} {\bibfnamefont {K.}~\bibnamefont {Yamada}}, \bibinfo
  {author} {\bibfnamefont {J.~M.}\ \bibnamefont {Tranquada}}, \ and\ \bibinfo
  {author} {\bibfnamefont {L.~P.}\ \bibnamefont {Regnault}},\ }\href {\doibase
  10.1103/PhysRevB.70.104517} {\bibfield  {journal} {\bibinfo  {journal} {Phys.
  Rev. B}\ }\textbf {\bibinfo {volume} {70}},\ \bibinfo {pages} {104517}
  (\bibinfo {year} {2004})}\BibitemShut {NoStop}%
\bibitem [{\citenamefont {Doering}\ \emph {et~al.}(2011)\citenamefont
  {Doering}, \citenamefont {Chuang}, \citenamefont {Andresen}, \citenamefont
  {Chow}, \citenamefont {Contarato}, \citenamefont {Cummings}, \citenamefont
  {Domning}, \citenamefont {Joseph}, \citenamefont {Pepper}, \citenamefont
  {Smith}, \citenamefont {Zizka}, \citenamefont {Ford}, \citenamefont {Lee},
  \citenamefont {Weaver}, \citenamefont {Patthey}, \citenamefont {Weizeorick},
  \citenamefont {Hussain},\ and\ \citenamefont {Denes}}]{fccd_camera}%
  \BibitemOpen
  \bibfield  {author} {\bibinfo {author} {\bibfnamefont {D.}~\bibnamefont
  {Doering}}, \bibinfo {author} {\bibfnamefont {Y.-D.}\ \bibnamefont {Chuang}},
  \bibinfo {author} {\bibfnamefont {N.}~\bibnamefont {Andresen}}, \bibinfo
  {author} {\bibfnamefont {K.}~\bibnamefont {Chow}}, \bibinfo {author}
  {\bibfnamefont {D.}~\bibnamefont {Contarato}}, \bibinfo {author}
  {\bibfnamefont {C.}~\bibnamefont {Cummings}}, \bibinfo {author}
  {\bibfnamefont {E.}~\bibnamefont {Domning}}, \bibinfo {author} {\bibfnamefont
  {J.}~\bibnamefont {Joseph}}, \bibinfo {author} {\bibfnamefont {J.~S.}\
  \bibnamefont {Pepper}}, \bibinfo {author} {\bibfnamefont {B.}~\bibnamefont
  {Smith}}, \bibinfo {author} {\bibfnamefont {G.}~\bibnamefont {Zizka}},
  \bibinfo {author} {\bibfnamefont {C.}~\bibnamefont {Ford}}, \bibinfo {author}
  {\bibfnamefont {W.~S.}\ \bibnamefont {Lee}}, \bibinfo {author} {\bibfnamefont
  {M.}~\bibnamefont {Weaver}}, \bibinfo {author} {\bibfnamefont
  {L.}~\bibnamefont {Patthey}}, \bibinfo {author} {\bibfnamefont
  {J.}~\bibnamefont {Weizeorick}}, \bibinfo {author} {\bibfnamefont
  {Z.}~\bibnamefont {Hussain}}, \ and\ \bibinfo {author} {\bibfnamefont
  {P.}~\bibnamefont {Denes}},\ }\href
  {http://scitation.aip.org/content/aip/journal/rsi/82/7/10.1063/1.3609862}
  {\bibfield  {journal} {\bibinfo  {journal} {Review of Scientific
  Instruments}\ }\textbf {\bibinfo {volume} {82}},\ \bibinfo {eid} {073303}
  (\bibinfo {year} {2011})}\BibitemShut {NoStop}%
\bibitem [{HKL()}]{HKL_notation}%
  \BibitemOpen
  \href@noop {} {}\bibinfo {note} {Negative $H$ denotes the CDW Bragg peak with
  smaller (rather than larger) incident x-ray angle.}\BibitemShut {Stop}%
\bibitem [{\citenamefont {Pitney}\ \emph {et~al.}(2000)\citenamefont {Pitney},
  \citenamefont {Robinson}, \citenamefont {Vartaniants}, \citenamefont
  {Appleton},\ and\ \citenamefont {Flynn}}]{Pitney2000}%
  \BibitemOpen
  \bibfield  {author} {\bibinfo {author} {\bibfnamefont {J.~A.}\ \bibnamefont
  {Pitney}}, \bibinfo {author} {\bibfnamefont {I.~K.}\ \bibnamefont
  {Robinson}}, \bibinfo {author} {\bibfnamefont {I.~A.}\ \bibnamefont
  {Vartaniants}}, \bibinfo {author} {\bibfnamefont {R.}~\bibnamefont
  {Appleton}}, \ and\ \bibinfo {author} {\bibfnamefont {C.~P.}\ \bibnamefont
  {Flynn}},\ }\href {\doibase 10.1103/PhysRevB.62.13084} {\bibfield  {journal}
  {\bibinfo  {journal} {Phys. Rev. B}\ }\textbf {\bibinfo {volume} {62}},\
  \bibinfo {pages} {13084} (\bibinfo {year} {2000})}\BibitemShut {NoStop}%
\bibitem [{\citenamefont {Li}\ \emph {et~al.}(2007)\citenamefont {Li},
  \citenamefont {H\"ucker}, \citenamefont {Gu}, \citenamefont {Tsvelik},\ and\
  \citenamefont {Tranquada}}]{Li2007}%
  \BibitemOpen
  \bibfield  {author} {\bibinfo {author} {\bibfnamefont {Q.}~\bibnamefont
  {Li}}, \bibinfo {author} {\bibfnamefont {M.}~\bibnamefont {H\"ucker}},
  \bibinfo {author} {\bibfnamefont {G.~D.}\ \bibnamefont {Gu}}, \bibinfo
  {author} {\bibfnamefont {A.~M.}\ \bibnamefont {Tsvelik}}, \ and\ \bibinfo
  {author} {\bibfnamefont {J.~M.}\ \bibnamefont {Tranquada}},\ }\href {\doibase
  10.1103/PhysRevLett.99.067001} {\bibfield  {journal} {\bibinfo  {journal}
  {Phys. Rev. Lett.}\ }\textbf {\bibinfo {volume} {99}},\ \bibinfo {pages}
  {067001} (\bibinfo {year} {2007})}\BibitemShut {NoStop}%
\bibitem [{\citenamefont {H\"ucker}\ \emph {et~al.}(2013)\citenamefont
  {H\"ucker}, \citenamefont {v.~Zimmermann}, \citenamefont {Xu}, \citenamefont
  {Wen}, \citenamefont {Gu},\ and\ \citenamefont {Tranquada}}]{Hucker2013}%
  \BibitemOpen
  \bibfield  {author} {\bibinfo {author} {\bibfnamefont {M.}~\bibnamefont
  {H\"ucker}}, \bibinfo {author} {\bibfnamefont {M.}~\bibnamefont
  {v.~Zimmermann}}, \bibinfo {author} {\bibfnamefont {Z.~J.}\ \bibnamefont
  {Xu}}, \bibinfo {author} {\bibfnamefont {J.~S.}\ \bibnamefont {Wen}},
  \bibinfo {author} {\bibfnamefont {G.~D.}\ \bibnamefont {Gu}}, \ and\ \bibinfo
  {author} {\bibfnamefont {J.~M.}\ \bibnamefont {Tranquada}},\ }\href {\doibase
  10.1103/PhysRevB.87.014501} {\bibfield  {journal} {\bibinfo  {journal} {Phys.
  Rev. B}\ }\textbf {\bibinfo {volume} {87}},\ \bibinfo {pages} {014501}
  (\bibinfo {year} {2013})}\BibitemShut {NoStop}%
\bibitem [{\citenamefont {Sutton}(2008)}]{Sutton2008}%
  \BibitemOpen
  \bibfield  {author} {\bibinfo {author} {\bibfnamefont {M.}~\bibnamefont
  {Sutton}},\ }\href {\doibase http://dx.doi.org/10.1016/j.crhy.2007.04.008}
  {\bibfield  {journal} {\bibinfo  {journal} {Comptes Rendus Physique}\
  }\textbf {\bibinfo {volume} {9}},\ \bibinfo {pages} {657 } (\bibinfo {year}
  {2008})}\BibitemShut {NoStop}%
\bibitem [{\citenamefont {Shpyrko}(2014)}]{Shpyrko2014}%
  \BibitemOpen
  \bibfield  {author} {\bibinfo {author} {\bibfnamefont {O.~G.}\ \bibnamefont
  {Shpyrko}},\ }\href {\doibase 10.1107/S1600577514018232} {\bibfield
  {journal} {\bibinfo  {journal} {Journal of Synchrotron Radiation}\ }\textbf
  {\bibinfo {volume} {21}},\ \bibinfo {pages} {1057} (\bibinfo {year}
  {2014})}\BibitemShut {NoStop}%
\bibitem [{Note1()}]{Note1}%
  \BibitemOpen
  \bibinfo {note} {The details on the calculations are available in the
  Supplementary Materials of Ref.~\cite {Chen2016}.}\BibitemShut {Stop}%
\bibitem [{Note2()}]{Note2}%
  \BibitemOpen
  \bibinfo {note} {See the Supplemental Material at XXX for further details on
  the temperature scaling of the speckle contrast factor}\BibitemShut {NoStop}%
\bibitem [{\citenamefont {Hoffman}\ \emph {et~al.}(2002)\citenamefont
  {Hoffman}, \citenamefont {Hudson}, \citenamefont {Lang}, \citenamefont
  {Madhavan}, \citenamefont {Eisaki}, \citenamefont {Uchida},\ and\
  \citenamefont {Davis}}]{Hoffman2002}%
  \BibitemOpen
  \bibfield  {author} {\bibinfo {author} {\bibfnamefont {J.}~\bibnamefont
  {Hoffman}}, \bibinfo {author} {\bibfnamefont {E.}~\bibnamefont {Hudson}},
  \bibinfo {author} {\bibfnamefont {K.}~\bibnamefont {Lang}}, \bibinfo {author}
  {\bibfnamefont {V.}~\bibnamefont {Madhavan}}, \bibinfo {author}
  {\bibfnamefont {H.}~\bibnamefont {Eisaki}}, \bibinfo {author} {\bibfnamefont
  {S.}~\bibnamefont {Uchida}}, \ and\ \bibinfo {author} {\bibfnamefont
  {J.}~\bibnamefont {Davis}},\ }\href@noop {} {\bibfield  {journal} {\bibinfo
  {journal} {Science}\ }\textbf {\bibinfo {volume} {295}},\ \bibinfo {pages}
  {466} (\bibinfo {year} {2002})}\BibitemShut {NoStop}%
\bibitem [{\citenamefont {Hanaguri}\ \emph {et~al.}(2004)\citenamefont
  {Hanaguri}, \citenamefont {Lupien}, \citenamefont {Kohsaka}, \citenamefont
  {Lee}, \citenamefont {Azuma}, \citenamefont {Takano}, \citenamefont
  {Takagi},\ and\ \citenamefont {Davis}}]{Hanaguri2004}%
  \BibitemOpen
  \bibfield  {author} {\bibinfo {author} {\bibfnamefont {T.}~\bibnamefont
  {Hanaguri}}, \bibinfo {author} {\bibfnamefont {C.}~\bibnamefont {Lupien}},
  \bibinfo {author} {\bibfnamefont {Y.}~\bibnamefont {Kohsaka}}, \bibinfo
  {author} {\bibfnamefont {D.-H.}\ \bibnamefont {Lee}}, \bibinfo {author}
  {\bibfnamefont {M.}~\bibnamefont {Azuma}}, \bibinfo {author} {\bibfnamefont
  {M.}~\bibnamefont {Takano}}, \bibinfo {author} {\bibfnamefont
  {H.}~\bibnamefont {Takagi}}, \ and\ \bibinfo {author} {\bibfnamefont
  {J.}~\bibnamefont {Davis}},\ }\href@noop {} {\bibfield  {journal} {\bibinfo
  {journal} {Nature}\ }\textbf {\bibinfo {volume} {430}},\ \bibinfo {pages}
  {1001} (\bibinfo {year} {2004})}\BibitemShut {NoStop}%
\bibitem [{\citenamefont {Parker}\ \emph {et~al.}(2010)\citenamefont {Parker},
  \citenamefont {Aynajian}, \citenamefont {da~Silva~Neto}, \citenamefont
  {Pushp}, \citenamefont {Ono}, \citenamefont {Wen}, \citenamefont {Xu},
  \citenamefont {Gu},\ and\ \citenamefont {Yazdani}}]{Parker2010}%
  \BibitemOpen
  \bibfield  {author} {\bibinfo {author} {\bibfnamefont {C.~V.}\ \bibnamefont
  {Parker}}, \bibinfo {author} {\bibfnamefont {P.}~\bibnamefont {Aynajian}},
  \bibinfo {author} {\bibfnamefont {E.~H.}\ \bibnamefont {da~Silva~Neto}},
  \bibinfo {author} {\bibfnamefont {A.}~\bibnamefont {Pushp}}, \bibinfo
  {author} {\bibfnamefont {S.}~\bibnamefont {Ono}}, \bibinfo {author}
  {\bibfnamefont {J.}~\bibnamefont {Wen}}, \bibinfo {author} {\bibfnamefont
  {Z.}~\bibnamefont {Xu}}, \bibinfo {author} {\bibfnamefont {G.}~\bibnamefont
  {Gu}}, \ and\ \bibinfo {author} {\bibfnamefont {A.}~\bibnamefont {Yazdani}},\
  }\href {http://dx.doi.org/10.1038/nature09597} {\bibfield  {journal}
  {\bibinfo  {journal} {Nature}\ }\textbf {\bibinfo {volume} {468}},\ \bibinfo
  {pages} {677} (\bibinfo {year} {2010})}\BibitemShut {NoStop}%
\bibitem [{\citenamefont {da~Silva~Neto}\ \emph {et~al.}(2015)\citenamefont
  {da~Silva~Neto}, \citenamefont {Comin}, \citenamefont {He}, \citenamefont
  {Sutarto}, \citenamefont {Jiang}, \citenamefont {Greene}, \citenamefont
  {Sawatzky},\ and\ \citenamefont {Damascelli}}]{daSilvaNeto2015}%
  \BibitemOpen
  \bibfield  {author} {\bibinfo {author} {\bibfnamefont {E.~H.}\ \bibnamefont
  {da~Silva~Neto}}, \bibinfo {author} {\bibfnamefont {R.}~\bibnamefont
  {Comin}}, \bibinfo {author} {\bibfnamefont {F.}~\bibnamefont {He}}, \bibinfo
  {author} {\bibfnamefont {R.}~\bibnamefont {Sutarto}}, \bibinfo {author}
  {\bibfnamefont {Y.}~\bibnamefont {Jiang}}, \bibinfo {author} {\bibfnamefont
  {R.~L.}\ \bibnamefont {Greene}}, \bibinfo {author} {\bibfnamefont {G.~A.}\
  \bibnamefont {Sawatzky}}, \ and\ \bibinfo {author} {\bibfnamefont
  {A.}~\bibnamefont {Damascelli}},\ }\href@noop {} {\bibfield  {journal}
  {\bibinfo  {journal} {Science}\ }\textbf {\bibinfo {volume} {347}},\ \bibinfo
  {pages} {282} (\bibinfo {year} {2015})}\BibitemShut {NoStop}%
\bibitem [{\citenamefont {Valla}\ \emph {et~al.}(2006)\citenamefont {Valla},
  \citenamefont {Fedorov}, \citenamefont {Lee}, \citenamefont {Davis},\ and\
  \citenamefont {Gu}}]{Valla2006}%
  \BibitemOpen
  \bibfield  {author} {\bibinfo {author} {\bibfnamefont {T.}~\bibnamefont
  {Valla}}, \bibinfo {author} {\bibfnamefont {A.~V.}\ \bibnamefont {Fedorov}},
  \bibinfo {author} {\bibfnamefont {J.}~\bibnamefont {Lee}}, \bibinfo {author}
  {\bibfnamefont {J.~C.}\ \bibnamefont {Davis}}, \ and\ \bibinfo {author}
  {\bibfnamefont {G.~D.}\ \bibnamefont {Gu}},\ }\href {\doibase
  10.1126/science.1134742} {\bibfield  {journal} {\bibinfo  {journal}
  {Science}\ }\textbf {\bibinfo {volume} {314}},\ \bibinfo {pages} {1914}
  (\bibinfo {year} {2006})}\BibitemShut {NoStop}%
\bibitem [{\citenamefont {Julien}\ \emph {et~al.}(2001)\citenamefont {Julien},
  \citenamefont {Campana}, \citenamefont {Rigamonti}, \citenamefont {Carretta},
  \citenamefont {Borsa}, \citenamefont {Kuhns}, \citenamefont {Reyes},
  \citenamefont {Moulton}, \citenamefont {Horvati\ifmmode~\acute{c}\else
  \'{c}\fi{}}, \citenamefont {Berthier}, \citenamefont {Vietkin},\ and\
  \citenamefont {Revcolevschi}}]{Julien2001}%
  \BibitemOpen
  \bibfield  {author} {\bibinfo {author} {\bibfnamefont {M.-H.}\ \bibnamefont
  {Julien}}, \bibinfo {author} {\bibfnamefont {A.}~\bibnamefont {Campana}},
  \bibinfo {author} {\bibfnamefont {A.}~\bibnamefont {Rigamonti}}, \bibinfo
  {author} {\bibfnamefont {P.}~\bibnamefont {Carretta}}, \bibinfo {author}
  {\bibfnamefont {F.}~\bibnamefont {Borsa}}, \bibinfo {author} {\bibfnamefont
  {P.}~\bibnamefont {Kuhns}}, \bibinfo {author} {\bibfnamefont {A.~P.}\
  \bibnamefont {Reyes}}, \bibinfo {author} {\bibfnamefont {W.~G.}\ \bibnamefont
  {Moulton}}, \bibinfo {author} {\bibfnamefont {M.}~\bibnamefont
  {Horvati\ifmmode~\acute{c}\else \'{c}\fi{}}}, \bibinfo {author}
  {\bibfnamefont {C.}~\bibnamefont {Berthier}}, \bibinfo {author}
  {\bibfnamefont {A.}~\bibnamefont {Vietkin}}, \ and\ \bibinfo {author}
  {\bibfnamefont {A.}~\bibnamefont {Revcolevschi}},\ }\href {\doibase
  10.1103/PhysRevB.63.144508} {\bibfield  {journal} {\bibinfo  {journal} {Phys.
  Rev. B}\ }\textbf {\bibinfo {volume} {63}},\ \bibinfo {pages} {144508}
  (\bibinfo {year} {2001})}\BibitemShut {NoStop}%
\bibitem [{\citenamefont {Pelc}\ \emph {et~al.}(2017)\citenamefont {Pelc},
  \citenamefont {Grafe}, \citenamefont {Gu},\ and\ \citenamefont
  {Po\ifmmode~\check{z}\else \v{z}\fi{}ek}}]{Pelc2017}%
  \BibitemOpen
  \bibfield  {author} {\bibinfo {author} {\bibfnamefont {D.}~\bibnamefont
  {Pelc}}, \bibinfo {author} {\bibfnamefont {H.-J.}\ \bibnamefont {Grafe}},
  \bibinfo {author} {\bibfnamefont {G.~D.}\ \bibnamefont {Gu}}, \ and\ \bibinfo
  {author} {\bibfnamefont {M.}~\bibnamefont {Po\ifmmode~\check{z}\else
  \v{z}\fi{}ek}},\ }\href {\doibase 10.1103/PhysRevB.95.054508} {\bibfield
  {journal} {\bibinfo  {journal} {Phys. Rev. B}\ }\textbf {\bibinfo {volume}
  {95}},\ \bibinfo {pages} {054508} (\bibinfo {year} {2017})}\BibitemShut
  {NoStop}%
\bibitem [{\citenamefont {{Wu}}\ \emph {et~al.}(2011)\citenamefont {{Wu}},
  \citenamefont {{Mayaffre}}, \citenamefont {{Kr{\"a}mer}}, \citenamefont
  {{Horvati{\'c}}}, \citenamefont {{Berthier}}, \citenamefont {{Hardy}},
  \citenamefont {{Liang}}, \citenamefont {{Bonn}},\ and\ \citenamefont
  {{Julien}}}]{Wu2011}%
  \BibitemOpen
  \bibfield  {author} {\bibinfo {author} {\bibfnamefont {T.}~\bibnamefont
  {{Wu}}}, \bibinfo {author} {\bibfnamefont {H.}~\bibnamefont {{Mayaffre}}},
  \bibinfo {author} {\bibfnamefont {S.}~\bibnamefont {{Kr{\"a}mer}}}, \bibinfo
  {author} {\bibfnamefont {M.}~\bibnamefont {{Horvati{\'c}}}}, \bibinfo
  {author} {\bibfnamefont {C.}~\bibnamefont {{Berthier}}}, \bibinfo {author}
  {\bibfnamefont {W.~N.}\ \bibnamefont {{Hardy}}}, \bibinfo {author}
  {\bibfnamefont {R.}~\bibnamefont {{Liang}}}, \bibinfo {author} {\bibfnamefont
  {D.~A.}\ \bibnamefont {{Bonn}}}, \ and\ \bibinfo {author} {\bibfnamefont
  {M.-H.}\ \bibnamefont {{Julien}}},\ }\href {\doibase 10.1038/nature10345}
  {\bibfield  {journal} {\bibinfo  {journal} {Nature}\ }\textbf {\bibinfo
  {volume} {477}},\ \bibinfo {pages} {191} (\bibinfo {year}
  {2011})}\BibitemShut {NoStop}%
\bibitem [{\citenamefont {Wu}\ \emph {et~al.}(2013)\citenamefont {Wu},
  \citenamefont {Mayaffre}, \citenamefont {Kr{\"a}mer}, \citenamefont
  {Horvati{\'c}}, \citenamefont {Berthier}, \citenamefont {Kuhns},
  \citenamefont {Reyes}, \citenamefont {Liang}, \citenamefont {Hardy},
  \citenamefont {Bonn} \emph {et~al.}}]{Wu2013_NMR}%
  \BibitemOpen
  \bibfield  {author} {\bibinfo {author} {\bibfnamefont {T.}~\bibnamefont
  {Wu}}, \bibinfo {author} {\bibfnamefont {H.}~\bibnamefont {Mayaffre}},
  \bibinfo {author} {\bibfnamefont {S.}~\bibnamefont {Kr{\"a}mer}}, \bibinfo
  {author} {\bibfnamefont {M.}~\bibnamefont {Horvati{\'c}}}, \bibinfo {author}
  {\bibfnamefont {C.}~\bibnamefont {Berthier}}, \bibinfo {author}
  {\bibfnamefont {P.~L.}\ \bibnamefont {Kuhns}}, \bibinfo {author}
  {\bibfnamefont {A.~P.}\ \bibnamefont {Reyes}}, \bibinfo {author}
  {\bibfnamefont {R.}~\bibnamefont {Liang}}, \bibinfo {author} {\bibfnamefont
  {W.}~\bibnamefont {Hardy}}, \bibinfo {author} {\bibfnamefont
  {D.}~\bibnamefont {Bonn}},  \emph {et~al.},\ }\href {\doibase
  10.1038/ncomms3113} {\bibfield  {journal} {\bibinfo  {journal} {Nature
  communications}\ }\textbf {\bibinfo {volume} {4}},\ \bibinfo {pages} {2113}
  (\bibinfo {year} {2013})}\BibitemShut {NoStop}%
\bibitem [{\citenamefont {Wu}\ \emph {et~al.}(2015)\citenamefont {Wu},
  \citenamefont {Mayaffre}, \citenamefont {Kr{\"a}mer}, \citenamefont
  {Horvati{\'c}}, \citenamefont {Berthier}, \citenamefont {Hardy},
  \citenamefont {Liang}, \citenamefont {Bonn},\ and\ \citenamefont
  {Julien}}]{Wu2015}%
  \BibitemOpen
  \bibfield  {author} {\bibinfo {author} {\bibfnamefont {T.}~\bibnamefont
  {Wu}}, \bibinfo {author} {\bibfnamefont {H.}~\bibnamefont {Mayaffre}},
  \bibinfo {author} {\bibfnamefont {S.}~\bibnamefont {Kr{\"a}mer}}, \bibinfo
  {author} {\bibfnamefont {M.}~\bibnamefont {Horvati{\'c}}}, \bibinfo {author}
  {\bibfnamefont {C.}~\bibnamefont {Berthier}}, \bibinfo {author}
  {\bibfnamefont {W.}~\bibnamefont {Hardy}}, \bibinfo {author} {\bibfnamefont
  {R.}~\bibnamefont {Liang}}, \bibinfo {author} {\bibfnamefont
  {D.}~\bibnamefont {Bonn}}, \ and\ \bibinfo {author} {\bibfnamefont {M.-H.}\
  \bibnamefont {Julien}},\ }\href {\doibase 10.1038/ncomms7438} {\bibfield
  {journal} {\bibinfo  {journal} {Nature communications}\ }\textbf {\bibinfo
  {volume} {6}},\ \bibinfo {pages} {6438} (\bibinfo {year} {2015})}\BibitemShut
  {NoStop}%
\bibitem [{\citenamefont {Shpyrko}\ \emph {et~al.}(2007)\citenamefont
  {Shpyrko}, \citenamefont {Isaacs}, \citenamefont {Logan}, \citenamefont
  {Feng}, \citenamefont {Aeppli}, \citenamefont {Jaramillo}, \citenamefont
  {Kim}, \citenamefont {Rosenbaum}, \citenamefont {Zschack}, \citenamefont
  {Sprung} \emph {et~al.}}]{Shpyrko2007}%
  \BibitemOpen
  \bibfield  {author} {\bibinfo {author} {\bibfnamefont {O.}~\bibnamefont
  {Shpyrko}}, \bibinfo {author} {\bibfnamefont {E.}~\bibnamefont {Isaacs}},
  \bibinfo {author} {\bibfnamefont {J.}~\bibnamefont {Logan}}, \bibinfo
  {author} {\bibfnamefont {Y.}~\bibnamefont {Feng}}, \bibinfo {author}
  {\bibfnamefont {G.}~\bibnamefont {Aeppli}}, \bibinfo {author} {\bibfnamefont
  {R.}~\bibnamefont {Jaramillo}}, \bibinfo {author} {\bibfnamefont
  {H.}~\bibnamefont {Kim}}, \bibinfo {author} {\bibfnamefont {T.}~\bibnamefont
  {Rosenbaum}}, \bibinfo {author} {\bibfnamefont {P.}~\bibnamefont {Zschack}},
  \bibinfo {author} {\bibfnamefont {M.}~\bibnamefont {Sprung}},  \emph
  {et~al.},\ }\href@noop {} {\bibfield  {journal} {\bibinfo  {journal}
  {Nature}\ }\textbf {\bibinfo {volume} {447}},\ \bibinfo {pages} {68}
  (\bibinfo {year} {2007})}\BibitemShut {NoStop}%
\bibitem [{\citenamefont {Su}\ \emph {et~al.}(2012)\citenamefont {Su},
  \citenamefont {Sandy}, \citenamefont {Mohanty}, \citenamefont {Shpyrko},\
  and\ \citenamefont {Sutton}}]{Su2012}%
  \BibitemOpen
  \bibfield  {author} {\bibinfo {author} {\bibfnamefont {J.-D.}\ \bibnamefont
  {Su}}, \bibinfo {author} {\bibfnamefont {A.~R.}\ \bibnamefont {Sandy}},
  \bibinfo {author} {\bibfnamefont {J.}~\bibnamefont {Mohanty}}, \bibinfo
  {author} {\bibfnamefont {O.~G.}\ \bibnamefont {Shpyrko}}, \ and\ \bibinfo
  {author} {\bibfnamefont {M.}~\bibnamefont {Sutton}},\ }\href {\doibase
  10.1103/PhysRevB.86.205105} {\bibfield  {journal} {\bibinfo  {journal} {Phys.
  Rev. B}\ }\textbf {\bibinfo {volume} {86}},\ \bibinfo {pages} {205105}
  (\bibinfo {year} {2012})}\BibitemShut {NoStop}%
\bibitem [{\citenamefont {Tranquada}\ \emph {et~al.}(2008)\citenamefont
  {Tranquada}, \citenamefont {Gu}, \citenamefont {H\"ucker}, \citenamefont
  {Jie}, \citenamefont {Kang}, \citenamefont {Klingeler}, \citenamefont {Li},
  \citenamefont {Tristan}, \citenamefont {Wen}, \citenamefont {Xu},
  \citenamefont {Xu}, \citenamefont {Zhou},\ and\ \citenamefont
  {v.~Zimmermann}}]{Tranquada2008}%
  \BibitemOpen
  \bibfield  {author} {\bibinfo {author} {\bibfnamefont {J.~M.}\ \bibnamefont
  {Tranquada}}, \bibinfo {author} {\bibfnamefont {G.~D.}\ \bibnamefont {Gu}},
  \bibinfo {author} {\bibfnamefont {M.}~\bibnamefont {H\"ucker}}, \bibinfo
  {author} {\bibfnamefont {Q.}~\bibnamefont {Jie}}, \bibinfo {author}
  {\bibfnamefont {H.-J.}\ \bibnamefont {Kang}}, \bibinfo {author}
  {\bibfnamefont {R.}~\bibnamefont {Klingeler}}, \bibinfo {author}
  {\bibfnamefont {Q.}~\bibnamefont {Li}}, \bibinfo {author} {\bibfnamefont
  {N.}~\bibnamefont {Tristan}}, \bibinfo {author} {\bibfnamefont {J.~S.}\
  \bibnamefont {Wen}}, \bibinfo {author} {\bibfnamefont {G.~Y.}\ \bibnamefont
  {Xu}}, \bibinfo {author} {\bibfnamefont {Z.~J.}\ \bibnamefont {Xu}}, \bibinfo
  {author} {\bibfnamefont {J.}~\bibnamefont {Zhou}}, \ and\ \bibinfo {author}
  {\bibfnamefont {M.}~\bibnamefont {v.~Zimmermann}},\ }\href {\doibase
  10.1103/PhysRevB.78.174529} {\bibfield  {journal} {\bibinfo  {journal} {Phys.
  Rev. B}\ }\textbf {\bibinfo {volume} {78}},\ \bibinfo {pages} {174529}
  (\bibinfo {year} {2008})}\BibitemShut {NoStop}%
\bibitem [{\citenamefont {Savici}\ \emph {et~al.}(2005)\citenamefont {Savici},
  \citenamefont {Fukaya}, \citenamefont {Gat-Malureanu}, \citenamefont {Ito},
  \citenamefont {Russo}, \citenamefont {Uemura}, \citenamefont {Wiebe},
  \citenamefont {Kyriakou}, \citenamefont {MacDougall}, \citenamefont {Rovers},
  \citenamefont {Luke}, \citenamefont {Kojima}, \citenamefont {Goto},
  \citenamefont {Uchida}, \citenamefont {Kadono}, \citenamefont {Yamada},
  \citenamefont {Tajima}, \citenamefont {Masui}, \citenamefont {Eisaki},
  \citenamefont {Kaneko}, \citenamefont {Greven},\ and\ \citenamefont
  {Gu}}]{Savici2005}%
  \BibitemOpen
  \bibfield  {author} {\bibinfo {author} {\bibfnamefont {A.~T.}\ \bibnamefont
  {Savici}}, \bibinfo {author} {\bibfnamefont {A.}~\bibnamefont {Fukaya}},
  \bibinfo {author} {\bibfnamefont {I.~M.}\ \bibnamefont {Gat-Malureanu}},
  \bibinfo {author} {\bibfnamefont {T.}~\bibnamefont {Ito}}, \bibinfo {author}
  {\bibfnamefont {P.~L.}\ \bibnamefont {Russo}}, \bibinfo {author}
  {\bibfnamefont {Y.~J.}\ \bibnamefont {Uemura}}, \bibinfo {author}
  {\bibfnamefont {C.~R.}\ \bibnamefont {Wiebe}}, \bibinfo {author}
  {\bibfnamefont {P.~P.}\ \bibnamefont {Kyriakou}}, \bibinfo {author}
  {\bibfnamefont {G.~J.}\ \bibnamefont {MacDougall}}, \bibinfo {author}
  {\bibfnamefont {M.~T.}\ \bibnamefont {Rovers}}, \bibinfo {author}
  {\bibfnamefont {G.~M.}\ \bibnamefont {Luke}}, \bibinfo {author}
  {\bibfnamefont {K.~M.}\ \bibnamefont {Kojima}}, \bibinfo {author}
  {\bibfnamefont {M.}~\bibnamefont {Goto}}, \bibinfo {author} {\bibfnamefont
  {S.}~\bibnamefont {Uchida}}, \bibinfo {author} {\bibfnamefont
  {R.}~\bibnamefont {Kadono}}, \bibinfo {author} {\bibfnamefont
  {K.}~\bibnamefont {Yamada}}, \bibinfo {author} {\bibfnamefont
  {S.}~\bibnamefont {Tajima}}, \bibinfo {author} {\bibfnamefont
  {T.}~\bibnamefont {Masui}}, \bibinfo {author} {\bibfnamefont
  {H.}~\bibnamefont {Eisaki}}, \bibinfo {author} {\bibfnamefont
  {N.}~\bibnamefont {Kaneko}}, \bibinfo {author} {\bibfnamefont
  {M.}~\bibnamefont {Greven}}, \ and\ \bibinfo {author} {\bibfnamefont {G.~D.}\
  \bibnamefont {Gu}},\ }\href {\doibase 10.1103/PhysRevLett.95.157001}
  {\bibfield  {journal} {\bibinfo  {journal} {Phys. Rev. Lett.}\ }\textbf
  {\bibinfo {volume} {95}},\ \bibinfo {pages} {157001} (\bibinfo {year}
  {2005})}\BibitemShut {NoStop}%
\bibitem [{\citenamefont {Hunt}\ \emph {et~al.}(2001)\citenamefont {Hunt},
  \citenamefont {Singer}, \citenamefont {Cederstr\"om},\ and\ \citenamefont
  {Imai}}]{Hunt2001}%
  \BibitemOpen
  \bibfield  {author} {\bibinfo {author} {\bibfnamefont {A.~W.}\ \bibnamefont
  {Hunt}}, \bibinfo {author} {\bibfnamefont {P.~M.}\ \bibnamefont {Singer}},
  \bibinfo {author} {\bibfnamefont {A.~F.}\ \bibnamefont {Cederstr\"om}}, \
  and\ \bibinfo {author} {\bibfnamefont {T.}~\bibnamefont {Imai}},\ }\href
  {\doibase 10.1103/PhysRevB.64.134525} {\bibfield  {journal} {\bibinfo
  {journal} {Phys. Rev. B}\ }\textbf {\bibinfo {volume} {64}},\ \bibinfo
  {pages} {134525} (\bibinfo {year} {2001})}\BibitemShut {NoStop}%
\bibitem [{\citenamefont {Bozin}\ \emph {et~al.}(2015)\citenamefont {Bozin},
  \citenamefont {Zhong}, \citenamefont {Knox}, \citenamefont {Gu},
  \citenamefont {Hill}, \citenamefont {Tranquada},\ and\ \citenamefont
  {Billinge}}]{Bozin2015}%
  \BibitemOpen
  \bibfield  {author} {\bibinfo {author} {\bibfnamefont {E.~S.}\ \bibnamefont
  {Bozin}}, \bibinfo {author} {\bibfnamefont {R.}~\bibnamefont {Zhong}},
  \bibinfo {author} {\bibfnamefont {K.~R.}\ \bibnamefont {Knox}}, \bibinfo
  {author} {\bibfnamefont {G.}~\bibnamefont {Gu}}, \bibinfo {author}
  {\bibfnamefont {J.~P.}\ \bibnamefont {Hill}}, \bibinfo {author}
  {\bibfnamefont {J.~M.}\ \bibnamefont {Tranquada}}, \ and\ \bibinfo {author}
  {\bibfnamefont {S.~J.~L.}\ \bibnamefont {Billinge}},\ }\href {\doibase
  10.1103/PhysRevB.91.054521} {\bibfield  {journal} {\bibinfo  {journal} {Phys.
  Rev. B}\ }\textbf {\bibinfo {volume} {91}},\ \bibinfo {pages} {054521}
  (\bibinfo {year} {2015})}\BibitemShut {NoStop}%
\bibitem [{\citenamefont {Reznik}\ \emph {et~al.}(2006)\citenamefont {Reznik},
  \citenamefont {Pintschovius}, \citenamefont {Ito}, \citenamefont {Iikubo},
  \citenamefont {Sato}, \citenamefont {Goka}, \citenamefont {Fujita},
  \citenamefont {Yamada}, \citenamefont {Gu},\ and\ \citenamefont
  {Tranquada}}]{Reznik2006}%
  \BibitemOpen
  \bibfield  {author} {\bibinfo {author} {\bibfnamefont {D.}~\bibnamefont
  {Reznik}}, \bibinfo {author} {\bibfnamefont {L.}~\bibnamefont
  {Pintschovius}}, \bibinfo {author} {\bibfnamefont {M.}~\bibnamefont {Ito}},
  \bibinfo {author} {\bibfnamefont {S.}~\bibnamefont {Iikubo}}, \bibinfo
  {author} {\bibfnamefont {M.}~\bibnamefont {Sato}}, \bibinfo {author}
  {\bibfnamefont {H.}~\bibnamefont {Goka}}, \bibinfo {author} {\bibfnamefont
  {M.}~\bibnamefont {Fujita}}, \bibinfo {author} {\bibfnamefont
  {K.}~\bibnamefont {Yamada}}, \bibinfo {author} {\bibfnamefont
  {G.}~\bibnamefont {Gu}}, \ and\ \bibinfo {author} {\bibfnamefont
  {J.}~\bibnamefont {Tranquada}},\ }\href@noop {} {\bibfield  {journal}
  {\bibinfo  {journal} {Nature}\ }\textbf {\bibinfo {volume} {440}},\ \bibinfo
  {pages} {1170} (\bibinfo {year} {2006})}\BibitemShut {NoStop}%
\bibitem [{\citenamefont {Berg}\ \emph {et~al.}(2007)\citenamefont {Berg},
  \citenamefont {Fradkin}, \citenamefont {Kim}, \citenamefont {Kivelson},
  \citenamefont {Oganesyan}, \citenamefont {Tranquada},\ and\ \citenamefont
  {Zhang}}]{Berg2007}%
  \BibitemOpen
  \bibfield  {author} {\bibinfo {author} {\bibfnamefont {E.}~\bibnamefont
  {Berg}}, \bibinfo {author} {\bibfnamefont {E.}~\bibnamefont {Fradkin}},
  \bibinfo {author} {\bibfnamefont {E.-A.}\ \bibnamefont {Kim}}, \bibinfo
  {author} {\bibfnamefont {S.~A.}\ \bibnamefont {Kivelson}}, \bibinfo {author}
  {\bibfnamefont {V.}~\bibnamefont {Oganesyan}}, \bibinfo {author}
  {\bibfnamefont {J.~M.}\ \bibnamefont {Tranquada}}, \ and\ \bibinfo {author}
  {\bibfnamefont {S.~C.}\ \bibnamefont {Zhang}},\ }\href {\doibase
  10.1103/PhysRevLett.99.127003} {\bibfield  {journal} {\bibinfo  {journal}
  {Phys. Rev. Lett.}\ }\textbf {\bibinfo {volume} {99}},\ \bibinfo {pages}
  {127003} (\bibinfo {year} {2007})}\BibitemShut {NoStop}%
\bibitem [{\citenamefont {Corboz}\ \emph {et~al.}(2014)\citenamefont {Corboz},
  \citenamefont {Rice},\ and\ \citenamefont {Troyer}}]{Corboz2014}%
  \BibitemOpen
  \bibfield  {author} {\bibinfo {author} {\bibfnamefont {P.}~\bibnamefont
  {Corboz}}, \bibinfo {author} {\bibfnamefont {T.~M.}\ \bibnamefont {Rice}}, \
  and\ \bibinfo {author} {\bibfnamefont {M.}~\bibnamefont {Troyer}},\ }\href
  {\doibase 10.1103/PhysRevLett.113.046402} {\bibfield  {journal} {\bibinfo
  {journal} {Phys. Rev. Lett.}\ }\textbf {\bibinfo {volume} {113}},\ \bibinfo
  {pages} {046402} (\bibinfo {year} {2014})}\BibitemShut {NoStop}%
\bibitem [{\citenamefont {Lee}(2014)}]{Lee2014}%
  \BibitemOpen
  \bibfield  {author} {\bibinfo {author} {\bibfnamefont {P.~A.}\ \bibnamefont
  {Lee}},\ }\href {\doibase 10.1103/PhysRevX.4.031017} {\bibfield  {journal}
  {\bibinfo  {journal} {Phys. Rev. X}\ }\textbf {\bibinfo {volume} {4}},\
  \bibinfo {pages} {031017} (\bibinfo {year} {2014})}\BibitemShut {NoStop}%
\bibitem [{\citenamefont {Fradkin}\ \emph {et~al.}(2015)\citenamefont
  {Fradkin}, \citenamefont {Kivelson},\ and\ \citenamefont
  {Tranquada}}]{Fradkin2015}%
  \BibitemOpen
  \bibfield  {author} {\bibinfo {author} {\bibfnamefont {E.}~\bibnamefont
  {Fradkin}}, \bibinfo {author} {\bibfnamefont {S.~A.}\ \bibnamefont
  {Kivelson}}, \ and\ \bibinfo {author} {\bibfnamefont {J.~M.}\ \bibnamefont
  {Tranquada}},\ }\href {\doibase 10.1103/RevModPhys.87.457} {\bibfield
  {journal} {\bibinfo  {journal} {Rev. Mod. Phys.}\ }\textbf {\bibinfo {volume}
  {87}},\ \bibinfo {pages} {457} (\bibinfo {year} {2015})}\BibitemShut
  {NoStop}%
\bibitem [{\citenamefont {Hamidian}\ \emph {et~al.}(2016)\citenamefont
  {Hamidian}, \citenamefont {Edkins}, \citenamefont {Joo}, \citenamefont
  {Kostin}, \citenamefont {Eisaki}, \citenamefont {Uchida}, \citenamefont
  {Lawler}, \citenamefont {Kim}, \citenamefont {Mackenzie}, \citenamefont
  {Fujita} \emph {et~al.}}]{Hamidian2016}%
  \BibitemOpen
  \bibfield  {author} {\bibinfo {author} {\bibfnamefont {M.}~\bibnamefont
  {Hamidian}}, \bibinfo {author} {\bibfnamefont {S.}~\bibnamefont {Edkins}},
  \bibinfo {author} {\bibfnamefont {S.~H.}\ \bibnamefont {Joo}}, \bibinfo
  {author} {\bibfnamefont {A.}~\bibnamefont {Kostin}}, \bibinfo {author}
  {\bibfnamefont {H.}~\bibnamefont {Eisaki}}, \bibinfo {author} {\bibfnamefont
  {S.}~\bibnamefont {Uchida}}, \bibinfo {author} {\bibfnamefont
  {M.}~\bibnamefont {Lawler}}, \bibinfo {author} {\bibfnamefont {E.-A.}\
  \bibnamefont {Kim}}, \bibinfo {author} {\bibfnamefont {A.}~\bibnamefont
  {Mackenzie}}, \bibinfo {author} {\bibfnamefont {K.}~\bibnamefont {Fujita}},
  \emph {et~al.},\ }\href@noop {} {\bibfield  {journal} {\bibinfo  {journal}
  {Nature}\ }\textbf {\bibinfo {volume} {532}},\ \bibinfo {pages} {343}
  (\bibinfo {year} {2016})}\BibitemShut {NoStop}%
\bibitem [{\citenamefont {Xu}\ \emph {et~al.}(2014)\citenamefont {Xu},
  \citenamefont {Stock}, \citenamefont {Chi}, \citenamefont {Kolesnikov},
  \citenamefont {Xu}, \citenamefont {Gu},\ and\ \citenamefont
  {Tranquada}}]{Xu2014}%
  \BibitemOpen
  \bibfield  {author} {\bibinfo {author} {\bibfnamefont {Z.}~\bibnamefont
  {Xu}}, \bibinfo {author} {\bibfnamefont {C.}~\bibnamefont {Stock}}, \bibinfo
  {author} {\bibfnamefont {S.}~\bibnamefont {Chi}}, \bibinfo {author}
  {\bibfnamefont {A.~I.}\ \bibnamefont {Kolesnikov}}, \bibinfo {author}
  {\bibfnamefont {G.}~\bibnamefont {Xu}}, \bibinfo {author} {\bibfnamefont
  {G.}~\bibnamefont {Gu}}, \ and\ \bibinfo {author} {\bibfnamefont {J.~M.}\
  \bibnamefont {Tranquada}},\ }\href {\doibase 10.1103/PhysRevLett.113.177002}
  {\bibfield  {journal} {\bibinfo  {journal} {Phys. Rev. Lett.}\ }\textbf
  {\bibinfo {volume} {113}},\ \bibinfo {pages} {177002} (\bibinfo {year}
  {2014})}\BibitemShut {NoStop}%
\end{thebibliography}%

\end{document}